# Production cross section measurements of radioactive isotopes by BigRIPS separator at RIKEN RI Beam Factory

H. Suzuki [a,*], T. Kubo [a,*], N. Fukuda [a], N. Inabe [a], D. Kameda [a], H. Takeda [a], K. Yoshida [a], K. Kusaka [a], Y. Yanagisawa [a], M. Ohtake [a], H. Sato [a], Y. Shimizu [a], H. Baba [a], M. Kurokawa [a], T. Ohnishi [a], K. Tanaka [a], O. B. Tarasov [b], D. Bazin [b], D. J. Morrissey [b], B. M. Sherrill [b], K. Ieki [c], D. Murai [c], N. Iwasa [d], A. Chiba [d], Y. Ohkoda [d], E. Ideguchi [e], S. Go [e], R. Yokoyama [e], T. Fujii [e], D. Nishimura [f], H. Nishibata [g], S. Momota [h], M. Lewitowicz [i], G. DeFrance [i], I. Celikovic [i], K. Steiger [j]

a *RIKEN Nishina Center, RIKEN, 2-1 Hirosawa, Wako, Saitama 351-0198, Japan*

b *National Superconducting Cyclotron Laboratory (NSCL), Michigan State University (MSU), 640 S. Shaw Lane, East Lansing, Michigan 48824-1321, USA*

c *Department of Physics, Rikkyo University, 3-34-1 Nishi-Ikebukuro, Toshima, Tokyo 171-8501, Japan*

d *Department of Physics, Tohoku University, 6-3 Aramaki, Aoba, Sendai 980-8578, Japan*

e *Center for Nuclear Study, University of Tokyo, 2-1 Hirosawa, Wako, Saitama 351-0198, Japan*

f *Faculty of Science and Technology, Tokyo University of Science, 2461 Yamazaki, Noda, Chiba 278-8510, Japan*

g *Department of Physics, Osaka University, 1-1 Machikaneyama, Toyonaka, Osaka 560-0034, Japan*

h *School of Environmental Science and Engineering, Kochi University of Technology, 185 Miyanokuchi, Tosayamada, Kami-city, Kochi 782-8502, Japan* i *Grand Accelerateur National d'Ions Lourds (GANIL), Bd Henri Becquerel, BP 55027, 14076 CAEN Cedex 05, France*

j *Physik-Department E12, Technische Universität München, James-Franck-Str. 1, D-85748 Garching, Germany*

**Abstract**

We have measured the production rates and production cross sections for a variety of radioactive isotopes which were produced from $^{124}$Xe, $^{48}$Ca, and $^{238}$U beams at an energy of 345 MeV/nucleon using the BigRIPS separator at the RIKEN Nishina Center RI Beam Factory (RIBF). Proton-rich isotopes with atomic numbers $Z$ = 40 to 52 and neutron-rich isotopes with $Z$ = 5 to 16 were produced by projectile fragmentation of the $^{124}$Xe and $^{48}$Ca beam on Be targets, respectively. Neutron-rich isotopes with $Z$ = 20 to 59 were produced by in-flight fission of the $^{238}$U beam, in which both Be and Pb were used as production targets. The measured production rates and production cross sections were compared with those of the LISE$^{++}$ calculations, and overall fairly good agreement has been obtained. Furthermore, in the measurements with the $^{124}$Xe beam, we have discovered four new isotopes on the proton-drip line, $^{85,86}$Ru and $^{81,82}$Mo, and obtained the clear evidence that $^{103}$Sb is particle unbound with an upper limit of 49 ns for the half-life. The measurements of projectile-fragment momentum distributions have been also performed with the $^{124}$Xe beam, in which the low-momentum tails of the distributions have been measured for the first time at the energy of 345 MeV/nucleon.

*Corresponding authors:
hsuzuki@ribf.riken.jp (Hiroshi Suzuki) and kubo@ribf.riken.jp (Toshiyuki Kubo)

## 1. Introduction

The production of radioactive isotope (RI) beams allows us to study exotic nuclei far from the stability, playing a major role in advancing nuclear physics. New interesting phenomena, such as neutron halos, neutron skins, and new magic numbers, have been discovered using RI beams and the region of accessible exotic nuclei has been expanding with the advancement of RI-beam facilities. Measurements of production cross sections for these RI beams are very important to carry out for the study of exotic nuclei. Production cross sections are crucial for designing experiments, where accurate estimates for the RI-beam intensities are required. Systematic measurements of production cross sections are also important to improve models and formulae which are used to predict cross sections.

Since the commissioning of the RIKEN RI Beam Factory (RIBF) [1] in March 2007, a wide variety of RI beams have been produced using the BigRIPS fragment separator [2-4], and used for various experiments with exotic nuclei such as secondary reaction measurements, decay measurements, and the search for new isotopes and new isomers. Neutron-rich exotic isotopes with atomic numbers $Z < 30$ were produced by projectile fragmentation [5] of $^{14}$N, $^{18}$O, $^{48}$Ca, and $^{70}$Zn beams at 345 MeV/nucleon. Proton-rich exotic isotopes with $Z \sim 40$ to 50 were produced by projectile fragmentation of a 345 MeV/nucleon $^{124}$Xe beam. A very wide range of neutron-rich exotic isotopes with $Z \sim 20$ to 70 were produced by in-flight fission [6,7] of a 345 MeV/nucleon $^{238}$U beam. The production cross sections of these exotic nuclei have been deduced from the measured production rates with the help of simulations.

The BigRIPS separator is characterized by large ion-optical acceptances, a two-stage tandem structure, and excellent in-flight particle identification. The horizontal and vertical angular acceptances are ±40 and ±50 mrad, respectively, and the maximum momentum acceptance is ±3%, allowing efficient collection of fragments produced by not only projectile fragmentation but also in-flight fission. The tandem structure allows two-stage isotope separation and delivery of tagged RI beams. These features have been well demonstrated in our experiments using in-flight fission of a $^{238}$U beam, in which we discovered a number of new isotopes and new isomers [8-10]. The details of the particle identification scheme are described in Ref. [11].

In this paper we present our measured production rates and cross sections from the projectile fragmentation of the $^{124}$Xe and $^{48}$Ca beams and from the in-flight fission of the $^{238}$U beam. The measured cross sections are compared with those calculated by the empirical cross-section formulae EPAX3.01 [12] and EPAX2.15 [13] for projectile fragmentation and with those predicted by the LISE$^{++}$ fission model [14] for in-flight fission. As for the measurements with the $^{124}$Xe beam, we also present our discovery of new isotopes and discuss the nuclear binding of some exotic nuclei located on the proton drip-line.

## 2. Experiment

Measurements were performed with $^{48}$Ca, $^{124}$Xe, and $^{238}$U beams that were accelerated to 345 MeV/nucleon by the cascade operation of the RIBF accelerator complex consisting of the linear accelerators, RILAC and RILAC-II and the four cyclotrons, RRC, fRC, IRC, and SRC [1]. Fig.1 shows a schematic layout of the BigRIPS separator. The experimental method was essentially the same as that described in Ref. [8,9]. The production targets were beryllium (Be) for measurements with the $^{48}$Ca, $^{124}$Xe, and $^{238}$U beams, but in addition, a lead (Pb) target was used for measurements with the $^{238}$U beam. The first stage of the BigRIPS separator collected and separated fragments, while the second stage served as a spectrometer for particle identification. An achromatic energy degrader was used in the first stage for the isotopic separation. If further purification was needed, another degrader was used in the second stage. The degrader material is aluminum (Al).

Particle identification was based on the TOF-$B\rho$-$\Delta E$ method, in which the time of flight (TOF), magnetic rigidity ($B\rho$), and energy loss ($\Delta E$) were measured to deduce the atomic number $Z$ and the mass-to-charge ratio $A/Q$ of fragments. The $Z$ value was deduced from the TOF and $\Delta E$, while the $B\rho$ and TOF were used to deduce the $A/Q$ value. The beam-line detectors used for the particle identification are shown in Fig. 1 along with the measurement scheme.

The TOF was measured between two thin plastic scintillators placed at the F3 and F7 foci, which are located at the beginning and the end of the second stage of the BigRIPS separator, respectively. The active area of the plastic scintillators is 100 or 120 mm in the horizontal direction and 100 mm in the vertical direction. We chose the thickness of the scintillators depending on the $Z$ values of fragments: it is typically 3, 1, and 0.2 mm for $Z < 20$, $20 < Z < 30$, and $Z > 30$, respectively. The length of the central flight-path between the plastic scintillators is 46.98 m. The $\Delta E$ was measured at F7 with a multi-sampling ionization chamber (MUSIC) [15] or a stack of silicon detectors. The MUSIC detector has an active area of 230 mm in diameter. Six energy-loss signals obtained from the MUSIC detector were averaged and used for the $\Delta E$ measurement. The silicon detectors were 0.35 mm thick, having an active area of 50 × 50 mm$^2$. We sometimes used plastic scintillators for the $\Delta E$ measurement in case of low-$Z$ fragments.

The $B\rho$ measurement was made by particle trajectory reconstruction not only in the first half of the second stage (F3 – F5) but also in the second half (F5 – F7), as shown in Fig.1. The fractional $B\rho$ deviation from a central value was reconstructed from measured particle trajectories in order to precisely determine the $B\rho$ value. For the trajectory reconstruction, the positions and angles of fragments were measured at the F3, F5, and F7 foci by using two sets of position-sensitive parallel plate avalanche counters (PPAC) [16,17] installed at the respective foci. The effective size of the PPACs is 150 or 240 mm in the horizontal direction and 150 mm in the vertical direction. First-order and high-order ion optical transfer maps were extracted from experimental data and were used for the trajectory reconstruction. The twofold $B\rho$ measurement is necessary to deduce the $A/Q$ value of fragments in combination with the TOF measurement, since fragments are slowed down at F5 due to the presence of the PPAC detectors and energy degrader. The PPAC detectors were also used for additional TOF measurements. The central $B\rho$ value of the BigRIPS separator is determined by using the magnetic fields of the dipole magnets measured by NMR probes and the central trajectory radii of the dipole magnets deduced from the magnetic field-map data [18]. The absolute $B\rho$ values of the fragments were calculated from the central $B\rho$ value thus obtained. The calibration of the TOF and $\Delta E$ measurements was also made by using this central $B\rho$ value. The details of the trajectory reconstruction are described in Ref. [11].

Particle identification is confirmed by measuring delayed $\gamma$ rays from known short-lived isomers. The observation of characteristic isomeric $\gamma$ rays allows unambiguous particle identification, a technique called isomer tagging.

The production cross sections were deduced from the measured production rates and the transmission efficiency in the BigRIPS separator. The production rate, defined as particles/s/pnA where 1 pnA (particle nano Ampere) = 6.24 × 10$^9$ particles/s, was deduced from the number of events of each isotope-peak in the $Z$ versus $A/Q$ particle identification plot and the measured primary-beam intensity after correcting for the detection efficiency of the detectors and the live-time of data acquisition system.

The transmission of each isotope including changes of charge state in the materials was estimated based on the simulations using the LISE$^{++}$ code [14], which were performed in Monte Carlo mode. The projectile-fragmentation mechanism is assumed to simulate measurements using the $^{48}$Ca and $^{124}$Xe beams, while the LISE$^{++}$ fission model [14] was used to simulate those using the $^{238}$U beams. We also investigated how secondary reactions in the target material increase production rates by using the LISE$^{++}$ code. (See the LISE$^{++}$ manual in Ref. [14] for the details of the secondary reaction effects).

The primary-beam intensity was monitored by measuring the rate of light charged particles recoiling out of the production target. A stack of three plastic scintillators were installed by the production target for this purpose. Because the calibration coefficient between the beam intensity and the rate of recoiling particles depended on the primary beam species, target material, target thickness, and size of beam spot, we calibrated the coefficient for every measurement.

## 3. Measurements with $^{124}$Xe beam and search for new isotopes

The following three measurements have been performed for proton-rich exotic isotopes with $Z = 40$ to 52 using a $^{124}$Xe beam at 345 MeV/nucleon:

- Systematic measurement of the production cross sections.
- Measurement of the fragment momentum distribution, especially for the low-momentum tail.
- Search for new isotopes near the proton-drip line.

Table I summarizes the BigRIPS settings used for the measurements. For the settings referred to as $^{100}$Sn, $^{101}$Sn, $^{102}$Sn, $^{103}$Sn, $^{104}$Sn, and $^{105}$Sn, the $B\rho$ values were tuned for the corresponding Sn isotope throughout the BigRIPS separator. The production cross sections have been measured for these proton-rich Sn isotopes including the double-magic nucleus $^{100}$Sn. Those of the isotopes transmitted together in each setting have been deduced as well. The settings referred to as $^{97}$Pd –3%, $^{97}$Pd, and $^{97}$Pd +3% were mainly aimed at investigating the low momentum tail in the momentum distribution. It is important to measure the momentum distribution for the production of the proton-rich isotopes, because the low-momentum tails of contaminant isotopes affect the purity of the isotope of interest, and the shapes of such tails vary greatly with the projectile energy. In the measurements with the settings referred to as $^{85}$Ru and $^{105}$Te, we have searched for proton-rich new isotopes in the regions with $Z$~42 to 44 and with $Z$~51 to 53, respectively. The $B\rho$ values were tuned for the corresponding isotopes throughout the separator. The momentum-tail measurements as well as the new isotope measurements also allowed us deduce the production cross sections for the transmitted isotopes.

The momentum acceptance of the BigRIPS separator, which is determined by the setting of the F1 slits, was set to ±2% (+1.5%/-2% in the $^{105}$Te setting), taking into consideration the total event rate. These acceptances are large enough to cover the momentum distribution of the produced fragments if the $B\rho$ setting is tuned for the peak momentum. The settings of the F2 and F7 slits were constrained by the amount of contaminants and by the limit of the count rate at F7. Our data acquisition system limited the count rate at F7 to 2 kHz. In the $^{105}$Te setting, the high-momentum side of F1 slits was closed to +1.5% to reduce the total event rate by cutting contaminant particles. The maximum intensity of the primary beam was typically 8 pnA and the beam was impinged on a 4.03-mm thick Be production target. The thickness of the plastic scintillators for the TOF measurement was 0.2 mm. The $\Delta E$ was measured at F7 using the MUSIC detector.

As an example, Fig. 3 shows the $Z$ versus $A/Q$ particle identification plot obtained with the $^{100}$Sn setting, in which events for $^{100}$Sn are clearly observed.

### 3.1 Momentum distributions

The transmission efficiency in the BigRIPS separator plays a major role in deducing the production cross sections from the measured production rates. The shape of the momentum distribution of fragments is particularly important to simulate the transmission efficiency, because the distribution exhibits the low-momentum tail and some isotopes were measured only in parts of their momentum tails. Fig. 3 shows our measured momentum distribution for $^{99}$Rh along with the simulations using the LISE$^{++}$ code (version 9.4.80). The low-momentum tail is clearly seen in the figure. The momentum measurement was performed by the trajectory reconstruction as described in the chapter 2.

In the LISE$^{++}$ code, the momentum distribution of projectile fragmentation is calculated from the "Universal parameterization" model developed by Tarasov [19]. This model is based on the convolution between a Gaussian distribution corresponding to the Goldhaber model [5] of fragmentation and an exponential attenuation arising from friction between the projectile spectator and participant, and can describe the width of the momentum distribution, the ratio of fragment and projectile velocity as well as the low-momentum tail. The numerical expressions of the model are given by Eqs. (5)–(7) in Ref. [19], in which the parameters $\sigma_{conv}$, $s$, and coef are used to fit the momentum distribution. The slope of the low-momentum tail is determined by the coef parameter.

The original values of $\sigma_{conv}$, $s$, and coef are 91.5, 0.1487, and 5.758, respectively, which are deduced by Tarasov [19] mainly using momentum spectra at energies lower than 100 MeV/nucleon. We searched the best coef value to reproduce the low-momentum tail at 345 MeV/nucleon, and found that a value of 1.9 gives a good description as shown in Fig. 3. We did not change the other two parameters from the original values. The obtained coef value is smaller than the original, indicating that the low-momentum tail falls off faster than the original parameterization. We also investigated the momentum distribution of other isotopes such as $^{89}$Zr, $^{92}$Mo, and $^{99}$Ag, and found that the value of 1.9 gives them a good description as well. In the present work the coef value of 1.9 was applied to all the isotopes produced using the $^{124}$Xe beam.

### 3.2 Production cross sections

Based on the transmission efficiency estimated from the LISE$^{++}$ (version 9.4.80) simulations, we have deduced the production cross sections for a number of isotopes produced by projectile fragmentation of the $^{124}$Xe beam. The LISE$^{++}$ simulations used the momentum distribution discussed in the previous section and the fragment angular distribution based on the Goldhaber model in which the dispersion effects due to the orbital deflection of the projectile are included [20]. The angular distribution was given by Eq. (2) in Ref. [20], in which the parameters of $\sigma_1$ = 90 MeV/c and $\sigma_2$ = 200 MeV/c were used. Note that the angular spreads of the projectile fragments are much smaller than the angular acceptances of the BigRIPS separator. The estimated transmission efficiency is ~70 to 80%, when the $B\rho$ settings are tuned for the isotope of interest throughout the BigRIPS separator, including the tuning for the peak momentum.

Fig. 4 shows the obtained production cross sections along with the predictions from the EPAX empirical formulae. The type of symbols shows which BigRIPS setting was used for the measurements. The filled symbols indicate that the distribution peak was located inside the slit opening at each focus. On the other hand, the open symbols indicate that the momentum distribution peak is located outside the slit opening at F1, or that the distribution peak at F2 is located outside the slit opening. The deduced cross sections shown with the open symbols are more dependent on the transmission and the distribution including the low-momentum tails. However the measured data of the same isotope obtained from different settings are fairly consistent with each other. Even though some isotopes were accepted only due to their low-momentum tails, the deduced cross sections are consistent, indicating the reliability of our measurements and simulations. We estimate that our method for determining the cross sections has a systematic error of ~50% for all the measurements with the $^{124}$Xe beam. This systematic error results from the evaluation of the transmission efficiency and the determination of the beam intensity.

The solid and dashed lines in Fig. 4 show the predicted cross sections from the empirical cross-section formulae EPAX3.01 [12] and EPAX2.15 [13], respectively. The measured cross sections are fairly well reproduced by these EPAX formulae, although some isotopes show some systematic discrepancies. Fig. 5 shows the comparison between the measured production rates for the proton-rich Sn isotopes and the LISE$^{++}$ predictions using the EPAX formulae, revealing that the measurements are systematically smaller than the predictions. In the case of $^{100}$Sn, the measured production cross section is $(7.4\pm1.7) \times 10^{-10}$ mb, while the calculated values with the EPAX3.01 and EPAX2.15 formulae are $5.76 \times 10^{-9}$ mb and $7.43 \times 10^{-9}$ mb, respectively. Our measured cross section at 345 MeV/nucleon is almost one order of magnitude smaller than the calculated values with the EPAX formulae. The production cross section of $^{100}$Sn was measured to be $(5.8\pm2.1) \times 10^{-9}$ mb at 1.0 GeV/nucleon by Hinke *et al.* [21], which is not far from the calculated values with the EPAX formulae, implying energy dependence of the production cross sections. We also observe that the discrepancy between the experimental data and the calculated values becomes significant with increasing $Z$ number at the projectile energy of 345 MeV/nucleon.

Table 2 lists the measured production rates for some selected isotopes along with the deduced production cross sections, so that the actual production yields can be calculated. Those listed are the most proton-rich isotopes observed in this work. The isotopes that are not fully transmitted due to the slit opening at F2 are labeled in Table 2 and the estimated transmission loss is given in the footnote.

The effects of the secondary reactions in the production target were estimated from the LISE$^{++}$ simulations using the EPAX3.01 formula. As the isotopes become more proton-rich, the effects get more significant due to their low production rates. It has been found that secondary reaction effects increase the simulated production rates by a factor of ~15 to 25 % for very proton-rich isotopes such as $^{100}$Sn, $^{105}$Te and the new isotopes described in the next section.

The intensity of the $^{124}$Xe beam reached ~30 pnA in June 2012.

### 3.3 Search for new isotopes

We have performed a search for new isotopes using two different settings of the BigRIPS separator, each targeting new proton-rich isotopes in regions with $Z$ ~42 to 44 and with $Z$ ~51 to 53. These settings are summarized in Table 1 as $^{85}$Ru and $^{105}$Te settings, respectively. The experimental methods and the data analyses are essentially the same as those used in our previous experiments, in which we identified a total of 47 new isotopes using in-flight fission of a 345 MeV/nucleon $^{238}$U beam [8,9].

Figs. 6 and 8 show the particle identification plots of $Z$ versus $A/Q$ for projectile fragments produced in the $^{85}$Ru setting and the $^{105}$Te setting, respectively. The solid lines indicate the limit of known isotopes. The relative root-mean-square (r.m.s.) $Z$ resolution and $A/Q$ resolution were typically 0.40% and 0.061% for the $^{85}$Ru setting, and 0.39% and 0.054% for the $^{105}$Te setting, respectively. These values were estimated from Zr ($Z = 40$) and Sn ($Z = 50$) isotopes, respectively. For the particle identification plot, we excluded inconsistent events by checking the phase space profiles of fragments, beam spot profiles, trajectories, and correlations made from the pulse-height signals and timing signals in the plastic scintillators, PPAC detectors, and MUSIC detector. We compared the $B\rho$ measurements before and after F5 to reject inconsistent events. This also allowed us to reject events whose charge state changed at F5. Furthermore, in order to confirm the particle identification, we observed delayed $\gamma$ rays emitted from the known isomers: those from $^{76}$Rb$^{m}$, $^{80}$Y$^{m}$, and $^{86}$Tc$^{m}$ for the $^{85}$Ru setting and those from $^{106}$Sb$^{m}$ for the $^{105}$Te setting.

In the measurement with the $^{85}$Ru setting, we have observed four proton-rich new isotopes $^{85,86}$Ru and $^{81,82}$Mo for the first time. The integrated beam dose and the total running time during the measurement were $3.73 \times 10^{15}$ particles and 21.75 h, respectively. As shown in Fig. 6 (b), these new isotopes are clearly seen for each corresponding value of $A/Q$, and are clearly separated from the neighboring isotopes. In this plot, fully-stripped $^{A}$Z isotopes and hydrogen-like $^{A-2}$Z isotopes are located very close to each other, because the $A/Q$ value is around 2.0. However, in the new isotope region, the latter isotopes are more neutron-deficient than the former ones, therefore such isotopes cannot be considered as contamination. The results of new isotopes are summarized in Table 3 along with their measured production cross sections. Fig. 7 shows a comparison of the measured cross sections of the new isotopes and neighboring known isotopes with the predictions from the EPAX3.01 and EPAX2.15 formulae. Although the predictions are somewhat overestimated for the new isotopes, the measured cross sections fairly well agree with the predictions, supporting our discovery of new isotopes.

As seen in Fig. 6, the isotopes $^{81}$Nb and $^{85}$Tc were not observed in our measurement, in contrast to the clear observation of other $N = Z - 1$ isotopes such as $^{79}$Zr, $^{83}$Mo, and $^{87}$Ru. These observations are consistent with a previous measurement by Janas *et al.* [22] which suggested that $^{81}$Nb and $^{85}$Tc are particle unbound. The experimental non-observation of $^{81}$Nb and $^{85}$Tc allows us to determine upper limits of the half-lives of these isotopes, based on their production yields expected from the systematics of the neighboring isotopes and their TOF values between F0 and F7. The TOF values of $^{81}$Nb and $^{85}$Tc are calculated to be 440 ns and 441 ns, respectively. The expected yields of $^{81}$Nb and $^{85}$Tc are 3400 counts and 1600 counts, respectively, if these isotopes are particle bound. Assuming the observation limit of one count and considering the decay in flight, we derived the upper limits of 38 ns and 42 ns for $^{81}$Nb and $^{85}$Tc, respectively. These values are a factor of ~2 shorter than those obtained by Janas *et al.* [22].

We also investigated the half-lives of the new isotopes $^{86}$Ru and $^{82}$Mo by comparing the observed yields with the extrapolated yields from the systematics of the neighboring isotopes. The observed and extrapolated yields are 6 and 8 counts for $^{86}$Ru and 35 and 24 counts for $^{82}$Mo, respectively. These two yields are comparable in both cases, suggesting that the decay loss in flight is small and that the half-lives of $^{86}$Ru and $^{82}$Mo are significantly longer than our TOF values: 438 ns for $^{86}$Ru and 437 ns for $^{82}$Mo.

In the measurement with the $^{105}$Te setting, we could not observe any new isotopes as shown in Fig. 8. However the production of $^{105}$Te via projectile fragmentation of a $^{124}$Xe beam was observed for the first time in our present work. The integrated beam dose and the total running time were $3.39 \times 10^{15}$ particles and 16.57 h, respectively. Isotope $^{105}$Te was previously produced by the fusion evaporation reaction of $^{50}$Cr($^{58}$Ni,3n) and identified through the observation of its $\alpha$ decay [23].

In the previous measurement by Rykaczewski, *et al.* [24], several events were observed for $^{103}$Sb after a TOF of ~1.5 μs. However the isotope $^{103}$Sb was not observed in our present measurement as seen in Fig. 8, indicating that $^{103}$Sb is particle unbound and the half-life is much shorter than our TOF value: 446 ns for $^{103}$Sb. Assuming the observation limit of one count and considering the yields expected relative to the neighboring isotopes, we derived an upper limit of 49 ns for the half-life of $^{103}$Sb. According to the yield systematics of the neighboring isotopes, 560 events of $^{103}$Sb are expected to be observed if this isotope is particle bound. Our present result provides clear evidence that $^{103}$Sb is particle unbound and that the proton drip-line of Sb isotopes is located here. The half-life of the neighboring isotope $^{104}$Sb is measured to be 0.44 s [25].

## 4. Measurements with $^{48}$Ca beam

Since 2008, a number of experiments have been carried out using neutron-rich RI beams produced by projectile fragmentation of a 345 MeV/nucleon $^{48}$Ca beam. Very neutron-rich RI beams have been used for reaction studies of exotic nuclei with the ZeroDegree Spectrometer [4]. In these experiments, the BigRIPS settings were optimized for the production of the RI beams, and the $B\rho$ values were tuned for them throughout the separator. We have measured the production rates in each experiment, which allowed us to deduce the production cross sections over a wide range of neutron-rich isotopes.

Fig. 9 shows the measured production cross sections along with predictions from the EPAX3.01 and EPAX2.15 formulae. The thickness of the Be production targets is shown by the symbols: The squares, circles, triangles, inverted triangles, diamonds, and pentagons in Fig. 9 indicate the measurements with the thickness of 5, 10, 15, 20, 30, and 40 mm, respectively. The transmission efficiency used in deducing the cross sections was estimated from the LISE$^{++}$ simulations, in which the momentum distribution and the angular distribution were calculated using the same methods as described in Section 3.2. As for the momentum distribution, we used the original coef parameter in the Tarasov's parameterization [19]. The behavior of the low-momentum tail did not much affect the transmission, because the BigRIPS separator was tuned for the peak of momentum distribution.

As seen in Fig. 9, the predictions from the EPAX2.15 formula are in better agreement with the measured cross sections than those from the EPAX3.01 formula. The EPAX3.01 formula is the latest parameterization to improve agreement especially for very neutron-rich fragments from medium-mass and heavy-mass projectiles [12], of which predicted cross sections were previously overestimated in the EPAX2 parameterizations. However the EPAX3.01 predictions tend to underestimate our measured cross sections as shown in Fig. 9. The revision seems to be too great in the present mass region.

Table 4 summarizes the measured production rates along with the deduced production cross sections, so that the actual production yields can be calculated. The intensity of the $^{48}$Ca beam has reached ~400 pnA in May 2012. The thickness of the Be targets and the effects of secondary reactions in the production target [14] are also listed in Table 4. Secondary reaction effects are given as the augmentation factor for the production rates, which was estimated from the LISE$^{++}$ simulations with the EPAX3.01 formula. The augmentation factor becomes larger as the target gets thicker and the isotopes get more neutron-rich. The augmentation factor amounts to ~3 for very neutron-rich isotopes such as $^{32}$Ne and $^{38}$Mg, according to the LISE$^{++}$ simulations. The secondary reaction effects reduce the cross sections, giving a favorable shift for the EPAX3.01 predictions as shown in Fig. 9.

## 5. Measurements with $^{238}$U beam

Since the RIBF became operational in March 2007, a number of experiments have been performed using a very wide variety of neutron-rich RI beams produced by in-flight fission of a 345 MeV/nucleon $^{238}$U beam. The experiments included measurements of $\beta$ and isomer decay, reaction measurements, such as $(p,p')$, Coulomb excitation, knockout reactions and secondary fragmentations, and a search for new isotopes and new isomers. The intensity of the $^{238}$U beam has reached ~10 pnA in autumn 2012, facilitating the measurements significantly. In these experiments we have measured the production rates of fission fragments and deduced the production cross sections.

Here we report on the measurements whose experimental conditions are listed in Table 5 and compare them with the predictions from the LISE$^{++}$ fission model [14]. The measurements have been performed for the $^{238}$U + Be (7 mm) and the $^{238}$U + Pb (1.5 mm) reactions at 345 MeV/nucleon, for which the $B\rho$ values were set to 7.249 and 6.992 Tm, respectively. In both measurements, no energy degraders were used in the BigRIPS separator, so that the transmission through the separator would be much less complex. This also allowed us to measure isotopes over a wide range of atomic numbers simultaneously. The experimental method is essentially the same as described in Ref. [8]. The uranium beam intensity was $4 \times 10^7$ particles/s on average. Fig. 10 shows the simulated $B\rho$ distribution of Ni isotopes produced in the $^{238}$U + Be reaction along with the $B\rho$ window of the BigRIPS separator. Note that the $B\rho$ setting was chosen to select the high-momentum side of the distribution, so that the overall counting rate might not be too high for our data acquisition system.

Figs. 11 and 12 show the production rates measured with the Be and Pb targets, respectively, along with the predictions from the LISE$^{++}$ (version 8.4.1) simulations in which the LISE$^{++}$ fission model is employed. The simulations were made in the Monte Carlo mode. Furthermore we used version 8.4.1 for these simulations to maintain consistency, because other in-flight fission measurements published in Ref. [8,9] were simulated using the same version. Note that the production rates shown in the figures are for events corresponding to fully-stripped ions throughout the BigRIPS separator.

The reaction mechanism in the LISE$^{++}$ fission model depends on the production targets we used: The abrasion fission (AF) model is applied for a low-$Z$ target, while for a high-$Z$ target the Coulomb fission (CF) model is used in combination with the AF model to include the contribution of Coulomb excitation of the $^{238}$U nucleus. We refer to the latter as the AF + CF model. The production cross sections of fission fragments are calculated using these models.

In our LISE$^{++}$ simulations, the AF model was used for the $^{238}$U + Be reaction, while the AF + CF model was applied to the $^{238}$U + Pb reaction. The AF model relies on the so-called three excitation energy model in which three nuclei, $^{236}$U, $^{226}$Th, and $^{220}$Ra are chosen to represent all the fissile nuclei created in the abrasion-ablation stage, followed by fission fragment distributions which are calculated based on the semi-empirical model from Ref. [26]. The standard parameters of this model, such as excitation energies and production cross sections of the fissile nuclei, are those reproducing the experimental data from the $^{238}$U + Be reaction at 750 MeV/nucleon [6]. The cross-section data of neutron-rich isotopes with $Z$ = 20 to 46 are used to obtain the parameters [14]. The standard parameters of the AF model thus obtained are listed in Table 6. While using AF + CF model, the three representative fissile nuclei are chosen to be $^{238}$U, $^{230}$Th, $^{214}$Po, and their parameters are listed in Table 7. These parameters are determined so as to reproduce the experimental data from the $^{238}$U + Pb reaction at 1000 MeV/nucleon [27]. In this case the cross-section data of neutron-rich isotopes with $Z$ = 31 to 59 are used to obtain the parameters [14]. We employed these recommended fission parameters in our present LISE$^{++}$ simulations. The details are given in the LISE$^{++}$ manual in Ref. [14].

Overall our measured production rates from the $^{238}$U + Be reaction are well reproduced by the LISE$^{++}$ predictions based on the AF model except for the isotopes with $Z > 50$. A large discrepancy is seen in this region. The LISE$^{++}$ predictions are orders of magnitude smaller than the measured rates. Similar discrepancies were observed in a different experiment in which neutron-rich isotopes with $55 < Z < 65$ were produced in the $^{238}$U + Be reaction at 345 MeV/nucleon [28]. We speculate that this is because the standard parameters of the AF model were obtained from cross-section data in the region of $Z$ = 20 to 46. The parameterization of the AF model needs to be improved incorporating cross-section data in the region of higher $Z$. We investigated the effects of the secondary reactions in the target, because a thick Be target was used. According to the LISE$^{++}$ simulations with the EPAX3.01 formula, overall the production rates do not increase much, except for very neutron-rich isotopes for which the augmentation factor typically amounts to ~2 to 3.

Our measured production rates from the $^{238}$U + Pb reaction are fairly well reproduced by the LISE$^{++}$ predictions based on the AF + CF model.

## 6. Summary

In summary, we have measured the production rates and cross sections for a variety of isotopes produced from the $^{124}$Xe, $^{48}$Ca, and $^{238}$U beams at 345 MeV/u using the BigRIPS separator at RIKEN RIBF. The proton-rich isotopes with $Z$ = 40 to 52 were produced by projectile fragmentation of the $^{124}$Xe beam, during which we measured the momentum distributions of the fragments as well as the production rates and cross sections. We have found that the exponential tails at the low-momentum side fall off faster than the LISE$^{++}$ calculation with the original parameterization. The EPAX cross-section formulae agreed fairly well with the experimental production cross sections which were deduced from the production rates and the transmission calculated with the LISE$^{++}$ simulation including the corrections for such tails. Furthermore, we have identified four new isotopes on the proton-drip line, $^{85,86}$Ru and $^{81,82}$Mo, and obtained the clear evidence that $^{103}$Sb is particle unbound with an upper limit of 49 ns for the half-life.

The neutron-rich isotopes with $Z$ = 5 to 16 were produced by projectile fragmentation of the $^{48}$Ca beam on Be targets. The EPAX2.15 formula reproduced the experimental production cross sections fairly well. The neutron-rich isotopes with $Z$ = 20 to 59 were produced by in-flight fission of the $^{238}$U beam on Be and Pb targets. The corresponding production rates were compared with the LISE$^{++}$ calculations, in which the AF model was used for the $^{238}$U + Be case and the AF + CF model for the $^{238}$U + Pb case. In the former case, the LISE$^{++}$ predictions agreed well with the experimental results except for the isotopes with $Z > 50$. The improvement of the AF model is needed to reproduce the production cross sections in the region of Z > 50. In the $^{238}$U + Pb case, overall the LISE$^{++}$ predictions fairly well reproduced the experimental data.

**Acknowledgement**

The present measurements were carried out at the RI Beam Factory operated by RIKEN Nishina Center, RIKEN and CNS, University of Tokyo. The authors are grateful to the RIBF accelerator crew for providing the primary beams. The authors would like to thank Dr. Y. Yano, RIKEN Nishina Center, for his support and encouragement. T.K. is grateful to Dr. J. Stasko for his careful reading of the manuscript.

Table 1. Summary of the experimental conditions used for measurements with a 345 MeV/nucleon $^{124}$Xe beam.

| Setting | $^{100}$Sn | $^{101}$Sn | $^{102}$Sn | $^{103}$Sn | $^{104}$Sn | $^{105}$Sn |
|---|---|---|---|---|---|---|
| Target (mm) | Be 4.03 | Be 4.03 | Be 4.03 | Be 4.03 | Be 4.03 | Be 4.03 |
| $B\rho$ after target (Tm) 1) | 5.255 | 5.310 | 5.362 | 5.417 | 5.468 | 5.523 |
| Degrader at F1 (mm) | Al 2.85 | Al 2.85 | Al 2.85 | Al 2.85 | Al 2.85 | Al 2.85 |
| $B\rho$ after F1 (Tm) 1) | 4.622 | 4.675 | 4.728 | 4.784 | 4.836 | 4.892 |
| Degrader at F5 (mm) | Al 2.99 | Al 2.99 | Al 2.99 | Al 2.99 | Al 2.99 | Al 2.99 |
| $B\rho$ after F5 (Tm) 1) | 3.668 | 3.731 | 3.788 | 3.848 | 3.907 | 3.967 |
| F1 slit (mm) (Δp/p) 2) | ±42.8 (±2%) | ±42.8 (±2%) | ±42.8 (±2%) | ±42.8 (±2%) | ±42.8 (±2%) | ±42.8 (±2%) |
| F2 slit (mm) | ±3 | ±5 | ±5 | ±10 | ±10 | ±10 |
| F7 slit (mm) | -15/+8 | ±6 | ±10 | ±15 | ±18 | ±15 |
| Isotope tuned 3) | $^{100}$Sn | $^{101}$Sn | $^{102}$Sn | $^{103}$Sn | $^{104}$Sn | $^{105}$Sn |

| Setting | $^{97}$Pd –3% | $^{97}$Pd | $^{97}$Pd +3% | $^{85}$Ru | $^{105}$Te |
|---|---|---|---|---|---|
| Target (mm) | Be 4.03 | Be 4.03 | Be 4.03 | Be 4.03 | Be 4.03 |
| $B\rho$ (Tm) after target1) | 5.411 | 5.572 | 5.739 | 5.114 | 5.300 |
| Degrader at F1 (mm) | Al 2.85 | Al 2.85 | Al 2.85 | Al 2.85 | Al 2.85 |
| $B\rho$ after F1 (Tm) 1) | 4.807 | 5.003 | 5.197 | 4.566 | 4.631 |
| Degrader at F5 (mm) | Al 2.99 | Al 2.99 | Al 2.99 | Al 1.97 | Al 1.97 |
| $B\rho$ after F5 (Tm) 1) | 3.940 | 4.201 | 4.457 | 4.065 | 3.996 |
| F1 slit (mm) (Δp/p) 2) | ±42.8 (±2%) | ±42.8 (±2%) | ±42.8 (±2%) | ±42.8 (±2%) | -32.1/+42.8 (+1.5/-2.0%) |
| F2 slit (mm) | ±18 | ±18 | ±18 | ±20 | -15/+20 |
| F7 slit (mm) | ±120 | ±120 | ±120 | ±20 | ±10 |
| Isotope tuned 3) | $^{97}$Pd 4) | $^{97}$Pd | $^{97}$Pd 5) | $^{85}$Ru | $^{105}$Te |

1) The $B\rho$ values from the magnetic fields of the dipole magnets in the BigRIPS separator.
2) Corresponding momentum acceptance of the BigRIPS separator.
3) The $B\rho$ settings are tuned for the listed isotope throughout the BigRIPS separator unless noted.
4) The $B\rho$ settings from F1 to F7 are tuned for $^{97}$Pd, while the $B\rho$ value after target is set to be 3% lower than the peak value of the momentum distribution of $^{97}$Pd.
5) The $B\rho$ settings from F1 to F7 are tuned for $^{97}$Pd, while the $B\rho$ value after target is set to be 3% higher than the peak value of the momentum distribution of $^{97}$Pd.

Table 2. Measured production rates and production cross sections of some selected isotopes produced in the reaction $^{124}$Xe + Be (4.03mm) at 345 MeV/nucleon.

| Nuclide | Setting | Production rate (cps/pnA) [1)] | Production cross section (mb) |
|---|---|---|---|
| $^{105}$Te | $^{105}$Te | 0.00030 | $1.2 \times 10^{-9}$ |
| $^{106}$Te | $^{105}$Te | 0.026 | $1.3 \times 10^{-7}$ |
| $^{104}$Sb | $^{105}$Te | 0.066 | $3.5 \times 10^{-7}$ |
| $^{100}$Sn [2)] | $^{100}$Sn | 0.00011 | $7.4 \times 10^{-10}$ |
| $^{101}$Sn [3)] | $^{101}$Sn | 0.0078 | $4.0 \times 10^{-8}$ |
| $^{102}$Sn [3)] | $^{102}$Sn | 0.51 | $2.2 \times 10^{-6}$ |
| $^{103}$Sn | $^{103}$Sn | 19 | $7.7 \times 10^{-5}$ |
| $^{104}$Sn | $^{104}$Sn | 280 | $1.2 \times 10^{-3}$ |
| $^{105}$Sn | $^{105}$Sn | 4200 | $1.7 \times 10^{-2}$ |
| $^{106}$Sn | $^{105}$Sn | 25000 | $1.1 \times 10^{-1}$ |
| $^{99}$In [2)] | $^{100}$Sn | 0.040 | $2.2 \times 10^{-7}$ |
| $^{100}$In [3)] | $^{101}$Sn | 1.4 | $8.6 \times 10^{-6}$ |
| $^{101}$In | $^{102}$Sn | 42 | $2.3 \times 10^{-4}$ |
| $^{102}$In | $^{103}$Sn | 920 | $4.2 \times 10^{-3}$ |
| $^{98}$Cd | $^{97}$Pd-3% | 4.8 | $1.8 \times 10^{-5}$ |
| $^{99}$Cd | $^{97}$Pd-3% | 180 | $6.7 \times 10^{-4}$ |
| $^{100}$Cd | $^{97}$Pd | 1600 | $7.1 \times 10^{-3}$ |
| $^{96}$Ag | $^{97}$Pd-3% | 17 | $6.5 \times 10^{-5}$ |
| $^{97}$Ag | $^{97}$Pd-3% | 320 | $1.2 \times 10^{-3}$ |
| $^{98}$Ag | $^{97}$Pd | 4300 | $1.8 \times 10^{-2}$ |
| $^{94}$Pd | $^{97}$Pd-3% | 9.4 | $4.1 \times 10^{-5}$ |
| $^{95}$Pd | $^{97}$Pd-3% | 740 | $2.8 \times 10^{-3}$ |
| $^{96}$Pd | $^{97}$Pd | 8700 | $3.7 \times 10^{-2}$ |
| $^{94}$Rh | $^{97}$Pd | 12000 | $5.2 \times 10^{-2}$ |
| $^{95}$Rh | $^{97}$Pd | 150000 | $5.6 \times 10^{-1}$ |
| $^{92}$Ru | $^{97}$Pd | 7700 | $3.6 \times 10^{-2}$ |

1) cps represents counts/s, while 1 pnA (particle-nano-Ampere) = $6.24 \times 10^9$ particles/s.
2) The production yield is lost by ~30 % at F2 due to the narrow slit opening such as ±3 mm, which corresponds to the root-mean-square width (1$\sigma$ width) of the horizontal beam image of fragments.

3) The production yield is lost by ~10 % at F2 due to the narrow slit opening such as ±5 mm. In case of the $^{101}Sn$ setting, a few more percent is lost at F7 due to the slit opening of ±6 mm.

Table 3 List of new isotopes identified in the present work.

| | Counts | Production rate (cps/pnA) [1)] | Production cross section (mb) |
|---|---|---|---|
| $^{85}Ru$ | 1 | $4 \times 10^{-6}$ | $2 \times 10^{-11}$ |
| $^{86}Ru$ | 6 | $2 \times 10^{-5}$ | $1 \times 10^{-10}$ |
| $^{81}Mo$ | 1 | $4 \times 10^{-6}$ | $2 \times 10^{-11}$ |
| $^{82}Mo$ | 35 | $1.4 \times 10^{-4}$ | $6.3 \times 10^{-10}$ |

1) cps represents counts/s, while 1 pnA (particle-nano-Ampere) = $6.24 \times 10^9$ particles/s.

Table 4. Measured production rates of neutron-rich isotopes produced in the reaction $^{48}Ca$ + Be at 345 MeV/nucleon shown along with the deduced production cross sections and the thickness of Be targets. The estimated effects of secondary reactions in the target are also given as the augmentation factor for the production rates.

| Isotopes [1)] | Production rate (cps / 100 pnA) [2)] | Cross section (mb) | Be target (mm) | $B\rho$ after target (Tm) | Degrader at F1 (mm) | $B\rho$ after F1 (Tm) | Degrader at F5 (mm) | $B\rho$ after F5 (Tm) | F1 slits (mm) | F2 slits (mm) | Augmentation factor [3)] | Year |
|---|---|---|---|---|---|---|---|---|---|---|---|---|
| $^{19}B$ | $5.50 \times 10^{0}$ | $3.44 \times 10^{-6}$ | 30 | 9.435 | 20 | 8.978 | 8 | 8.778 | ±120 | -7/+10 | 1.91 | 2012 |
| $^{19}C$ | $2.17 \times 10^{4}$ | $1.33 \times 10^{-3}$ | 20 | 8.400 | 15 | 8.033 | 8 | 7.825 | ±43 | ±10 | 1.35 | 2008 |
| $^{19}C$ | $1.26 \times 10^{4}$ | $9.05 \times 10^{-4}$ | 30 | 8.116 | 20 | 7.594 | 8 | 7.337 | ±64 | ±5 | 1.56 | 2012 |
| $^{20}C$ | $4.14 \times 10^{3}$ | $1.83 \times 10^{-4}$ | 20 | 8.591 | 15 | 8.216 | --- | --- | ±120 | ±10 | 1.44 | 2008 |
| $^{20}C$ | $3.21 \times 10^{3}$ | $1.88 \times 10^{-4}$ | 20 | 8.591 | 15 | 8.212 | 8 | 7.967 | ±64 | ±20 | 1.44 | 2012 |
| $^{22}C$ | $5.29 \times 10^{0}$ | $2.19 \times 10^{-7}$ | 20 | 9.400 | 15 | 9.021 | --- | --- | ±120 | -20/+120 | 1.91 | 2008 |
| $^{22}C$ | $6.95 \times 10^{0}$ | $2.82 \times 10^{-7}$ | 20 | 9.400 | 15 | 9.017 | 8 | 8.767 | ±120 | -5/+10 | 1.91 | 2012 |
| $^{22}N$ | $4.00 \times 10^{3}$ | $1.88 \times 10^{-4}$ | 20 | 8.271 | 15 | 7.837 | 8 | 7.547 | ±64 | ±7 | 1.44 | 2012 |
| $^{22}O$ | $6.20 \times 10^{5}$ | $2.42 \times 10^{-2}$ | 15 | 7.409 | 15 | 6.936 | --- | --- | ±64 | ±3 | 1.21 | 2010 |
| $^{22}O$ | $2.93 \times 10^{5}$ | $3.07 \times 10^{-2}$ | 20 | 7.104 | 8 | 6.821 | --- | --- | ±21 | ±5 | 1.28 | 2012 |
| $^{24}O$ | $1.48 \times 10^{3}$ | $4.07 \times 10^{-5}$ | 15 | 8.106 | 15 | 7.633 | --- | --- | ±64 | -5/+6 | 1.32 | 2010 |
| $^{24}O$ | $1.18 \times 10^{2}$ | $1.83 \times 10^{-4}$ | 20 | 7.817 | 15 | 7.308 | --- | --- | ±2 | ±5 | 1.39 | 2012 |
| $^{24}O$ | $1.29 \times 10^{3}$ | $1.18 \times 10^{-4}$ | 15 | 7.909 | 20 | 7.222 | --- | --- | ±21 | ±8 | 1.32 | 2012 |
| $^{21}Ne$ | $1.97 \times 10^{7}$ | $7.04 \times 10^{1}$ | 10 | 5.409 | 6 | 5.158 | --- | --- | ±3 | -20/+10 | 1.14 | 2009 |
| $^{23}Ne$ | $8.00 \times 10^{7}$ | $5.79 \times 10^{0}$ | 40 | 5.409 | 6 | 5.118 | --- | --- | ±60 | ±20 | 1.50 | 2009 |
| $^{26}Ne$ | $4.40 \times 10^{6}$ | $1.30 \times 10^{-1}$ | 20 | 6.850 | 10 | 6.444 | --- | --- | ±64 | ±10 | 1.24 | 2009 |
| $^{28}Ne$ | $9.53 \times 10^{4}$ | $2.55 \times 10^{-3}$ | 15 | 7.599 | 15 | 6.995 | --- | --- | ±120 | -20/+8 | 1.26 | 2009 |
| $^{29}Ne$ | $1.84 \times 10^{3}$ | $7.04 \times 10^{-5}$ | 10 | 7.918 | 15 | 7.332 | --- | --- | ±120 | ±120 | 1.23 | 2008 |
| $^{29}Ne$ | $2.00 \times 10^{3}$ | $1.25 \times 10^{-4}$ | 15 | 7.706 | 12 | 7.210 | --- | --- | ±64 | ±5 | 1.35 | 2010 |
| $^{30}Ne$ | $5.50 \times 10^{2}$ | $1.38 \times 10^{-5}$ | 15 | 7.918 | 15 | 7.296 | --- | --- | ±120 | ±5 | 1.52 | 2008 |
| $^{30}Ne$ | $4.60 \times 10^{2}$ | $9.63 \times 10^{-6}$ | 15 | 8.149 | 15 | 7.555 | 5 | 7.294 | ±64 | ±5 | 1.52 | 2010 |
| $^{31}Ne$ | $1.25 \times 10^{1}$ | $2.45 \times 10^{-7}$ | 15 | 8.200 | 15 | 7.575 | --- | --- | ±120 | ±5 | 1.79 | 2008 |
| $^{32}Ne$ | $3.41 \times 10^{0}$ | $6.89 \times 10^{-8}$ | 20 | 8.400 | 15 | 7.779 | --- | --- | ±120 | ±5 | 2.81 | 2008 |
| $^{25}Mg$ | $2.00 \times 10^{8}$ | $7.14 \times 10^{0}$ | 10 | 5.500 | 8 | 5.110 | --- | --- | ±64 | ±20 | 1.14 | 2010 |
| $^{27}Mg$ | $2.37 \times 10^{8}$ | $5.84 \times 10^{0}$ | 10 | 6.000 | 8 | 5.614 | --- | --- | ±64 | ±20 | 1.09 | 2010 |
| $^{29}Mg$ | $6.00 \times 10^{5}$ | $5.17 \times 10^{-1}$ | 20 | 6.342 | 8 | 5.948 | --- | --- | ±2 | ±15 | 1.18 | 2010 |
| $^{32}Mg$ | $5.60 \times 10^{5}$ | $9.49 \times 10^{-3}$ | 20 | 6.807 | 15 | 5.982 | --- | --- | ±60 | ±15 | 1.27 | 2009 |
| $^{32}$Mg | $1.90 \times 10^{4}$ | $9.64 \times 10^{-3}$ | 10 | 7.320 | 8 | 6.947 | --- | --- | ±2 | ±15 | 1.13 | 2010 |
| $^{33}$Mg | $7.30 \times 10^{3}$ | $1.44 \times 10^{-3}$ | 15 | 7.269 | 10 | 6.769 | --- | --- | ±5 | ±5 | 1.26 | 2010 |
| $^{34}$Mg | $6.57 \times 10^{3}$ | $1.25 \times 10^{-4}$ | 15 | 7.392 | 8 | 6.993 | --- | --- | ±64 | ±4 | 1.36 | 2010 |
| $^{35}$Mg | $2.78 \times 10^{2}$ | $5.12 \times 10^{-6}$ | 15 | 7.812 | 10 | 7.329 | --- | --- | ±64 | ±4 | 1.54 | 2010 |
| $^{36}$Mg | $9.00 \times 10^{1}$ | $1.76 \times 10^{-6}$ | 15 | 7.986 | 10 | 7.497 | 5 | 7.176 | ±64 | ±5 | 1.84 | 2010 |
| $^{37}$Mg | $3.55 \times 10^{0}$ | $6.86 \times 10^{-8}$ | 15 | 8.206 | 10 | 7.719 | --- | --- | ±64 | ±12 | 2.43 | 2010 |
| $^{38}$Mg | $1.50 \times 10^{0}$ | $2.93 \times 10^{-8}$ | 15 | 8.436 | 8 | 8.049 | --- | --- | ±64 | ±15 | 3.54 | 2010 |
| $^{34}$Al | $2.60 \times 10^{5}$ | $2.12 \times 10^{-2}$ | 15 | 6.885 | 8 | 6.447 | 5 | 6.120 | ±64 | ±12 | 1.17 | 2010 |
| $^{41}$Al | $6.30 \times 10^{-1}$ | $1.09 \times 10^{-8}$ | 15 | 8.364 | 8 | 7.938 | 5 | 7.628 | ±64 | ±5 | 2.49 | 2010 |
| $^{39}$Si | $1.10 \times 10^{4}$ | $1.70 \times 10^{-4}$ | 15 | 7.353 | 10 | 6.764 | --- | --- | ±64 | ±4 | 1.33 | 2010 |
| $^{40}$Si | $2.06 \times 10^{3}$ | $5.97 \times 10^{-5}$ | 5 | 8.045 | 10 | 7.519 | --- | --- | ±60 | ±15 | 1.14 | 2008 |
| $^{40}$Si | $5.00 \times 10^{3}$ | $1.23 \times 10^{-4}$ | 15 | 7.672 | 10 | 7.098 | 5 | 6.751 | ±64 | -5/+3.5 | 1.44 | 2010 |
| $^{41}$Si | $3.50 \times 10^{2}$ | $5.60 \times 10^{-6}$ | 15 | 7.870 | 10 | 7.301 | --- | --- | ±64 | ±12 | 1.51 | 2010 |
| $^{42}$Si | $2.45 \times 10^{1}$ | $3.26 \times 10^{-7}$ | 20 | 7.686 | 10 | 7.070 | --- | --- | ±60 | ±15 | 1.60 | 2008 |
| $^{42}$Si | $2.38 \times 10^{1}$ | $3.88 \times 10^{-7}$ | 15 | 7.949 | 8 | 7.487 | 5 | 7.139 | ±64 | ±5 | 1.45 | 2010 |
| $^{40}$S | $5.30 \times 10^{5}$ | $1.46 \times 10^{-1}$ | 15 | 6.653 | 8 | 6.125 | 5 | 5.660 | ±2 | ±4 | 1.09 | 2010 |
| $^{42}$S | $1.40 \times 10^{5}$ | $4.13 \times 10^{-2}$ | 15 | 6.883 | 10 | 6.186 | 5 | 5.726 | ±2 | ±4 | 1.10 | 2010 |
| $^{44}$S | $3.00 \times 10^{4}$ | $1.91 \times 10^{-3}$ | 15 | 7.226 | 8 | 6.681 | 5 | 6.221 | ±21 | ±4 | 1.08 | 2010 |

1) The BigRIPS settings were optimized for the production of the isotopes in most cases and the $B\rho$ values were tuned for them throughout the separator. The slit settings at F5 and F7 do not affect the transmission of the isotopes.

2) cps represents counts/s, while 1 pnA (particle-nano-Ampere) = $6.24 \times 10^{9}$ particles/s.

3) The augmentation factor was estimated from the LISE$^{++}$ simulations with the EPAX3.01 formula. Note that the simulations were made for the exit of the target, because no significant difference was found between the target exit and BigRIPS exit.

Table 5. Summary of the experimental conditions for the measurements with a $^{238}$U beam at 345 MeV/nucleon.

| | $^{238}$U + Be | $^{238}$U + Pb |
|---|---|---|
| Target (mm) | Be 7 | Pb 1.5 |
| $B\rho$ (Tm) 1) | 7.249 | 6.992 |
| Degrader at F1 (mm) | no | no |
| Degrader at F5 (mm) | no | no |
| F1 slit (mm)<br>($\Delta p/p$) 2) | ±21<br>(±1%) | ±2<br>(±0.1%) |
| F2 slit (mm) 3) | ±30 | ±50 |
| F7 slit (mm) 3) | ±120 | ±120 |

1) The $B\rho$ values from the magnetic fields of the first dipole magnet in the BigRIPS separator. The rest of the sections have essentially the same $B\rho$ value because no energy degraders are used.
2) Corresponding momentum acceptance of the BigRIPS separator.
3) The slits at F2 and F7 do not affect the transmission because no energy degraders are used.

Table 6. Standard parameters of the LISE$^{++}$ fission model used for the $^{238}U$ + Be reaction (see the LISE$^{++}$ manual in Ref. [14] for the details).

| | Low | Middle | High |
|---|---|---|---|
| Fissile 1) | $^{236}U$ | $^{226}Th$ | $^{220}Ra$ |
| E [MeV] 2) | 23.5 | 100 | 250 |
| $\sigma$ [mb] 3) | 200 | 500 | 350 |

1) Fissile nuclei created in the abrasion-ablation stage.

2) Excitation energies of the fissile nuclei.

3) Production cross sections of the fissile nuclei.

Table 7. Standard parameters of the LISE++ fission model used for the $^{238}U$ + Pb reaction (see the LISE++ manual in Ref. [14] for the details).

| | Low | Middle | High |
|---|---|---|---|
| Fissile | $^{238}U$ | $^{230}Th$ | $^{214}Po$ |
| E [MeV] | 17.3 | 100 | 300 |
| σ [mb] | 2280 | 500 | 1300 |

See the footnote of Tale 6.

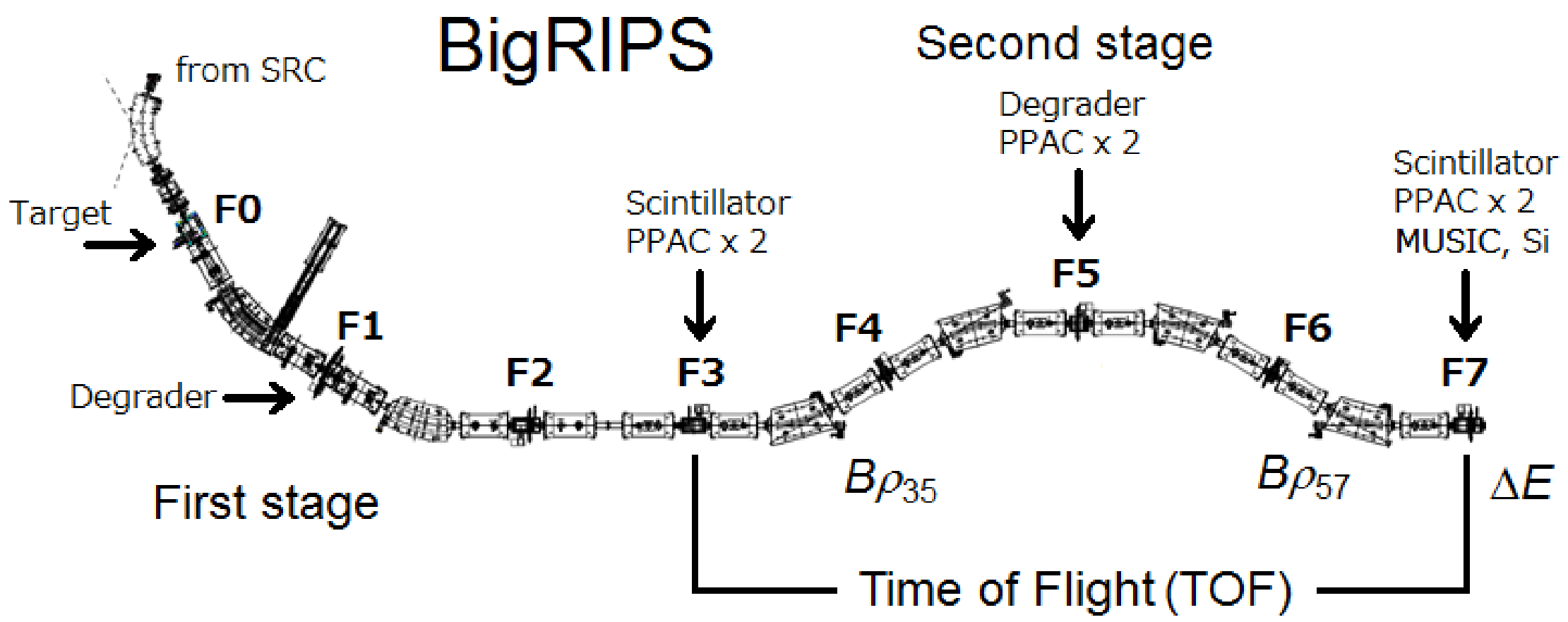

Fig. 1. Schematic layout of the BigRIPS separator. The label Fn indicates the positions of foci. The first stage of the BigRIPS separator includes components from the production target (F0) to F2, while the second stage spans F3 to F7. The foci at F3 and F7 are fully achromatic, while those at F1 and F5 are momentum dispersive. Energy degraders are placed at the dispersive foci. The positions of the beam-line detectors used for the particle identification are also shown along with the measurement scheme for time of flight (TOF), magnetic rigidity ($B\rho_{35}$ and $B\rho_{57}$), and energy loss ($\Delta E$). See text.

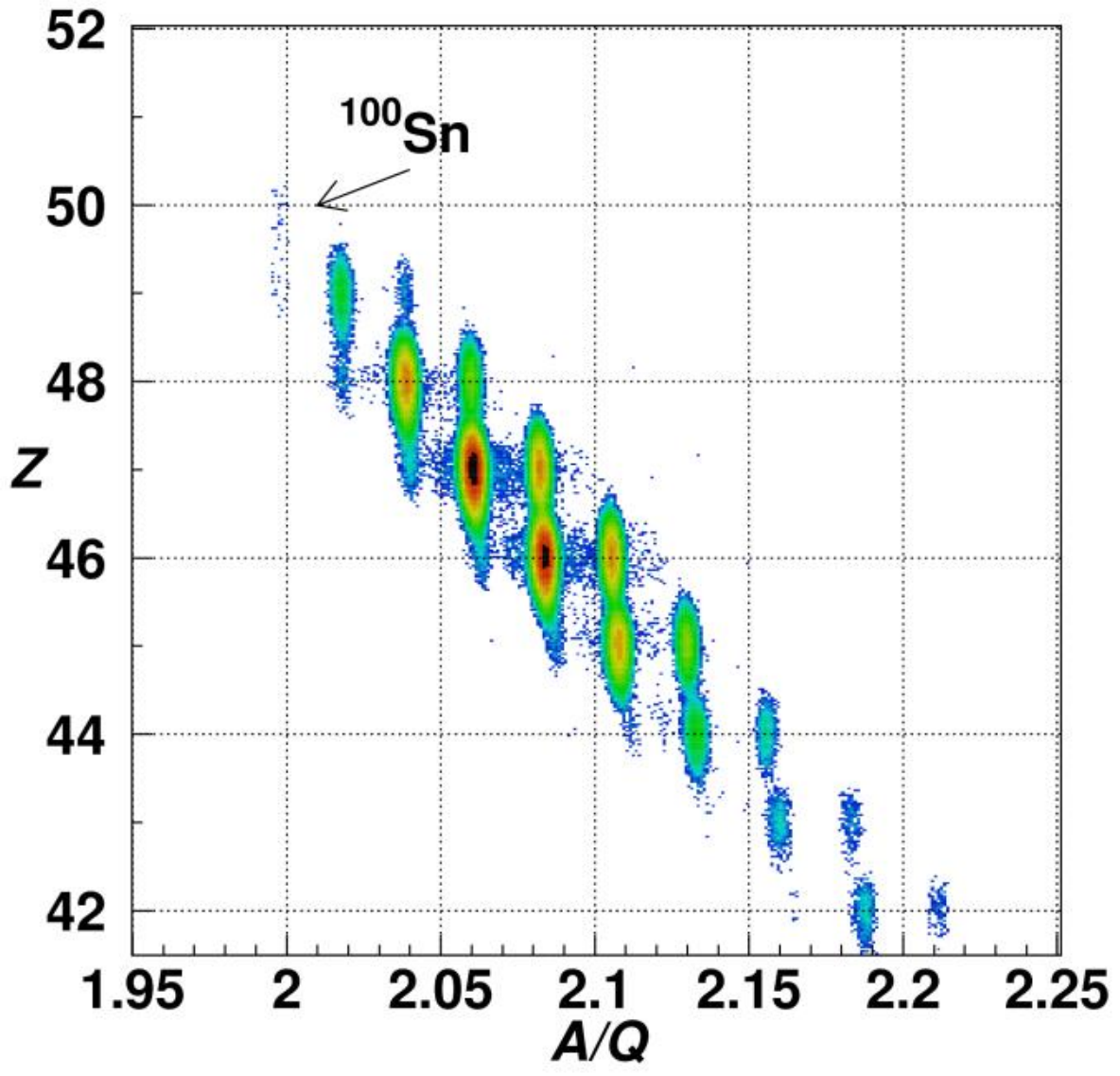

Fig. 2. $Z$ versus $A/Q$ particle identification plot for projectile fragments produced in the reaction $^{124}$Xe + Be (4.03 mm) at 345 MeV/nucleon. Experimental conditions are given as $^{100}$Sn setting in Table 1, in which the $B\rho$ settings are tuned for $^{100}$Sn throughout the BigRIPS separator. See text.

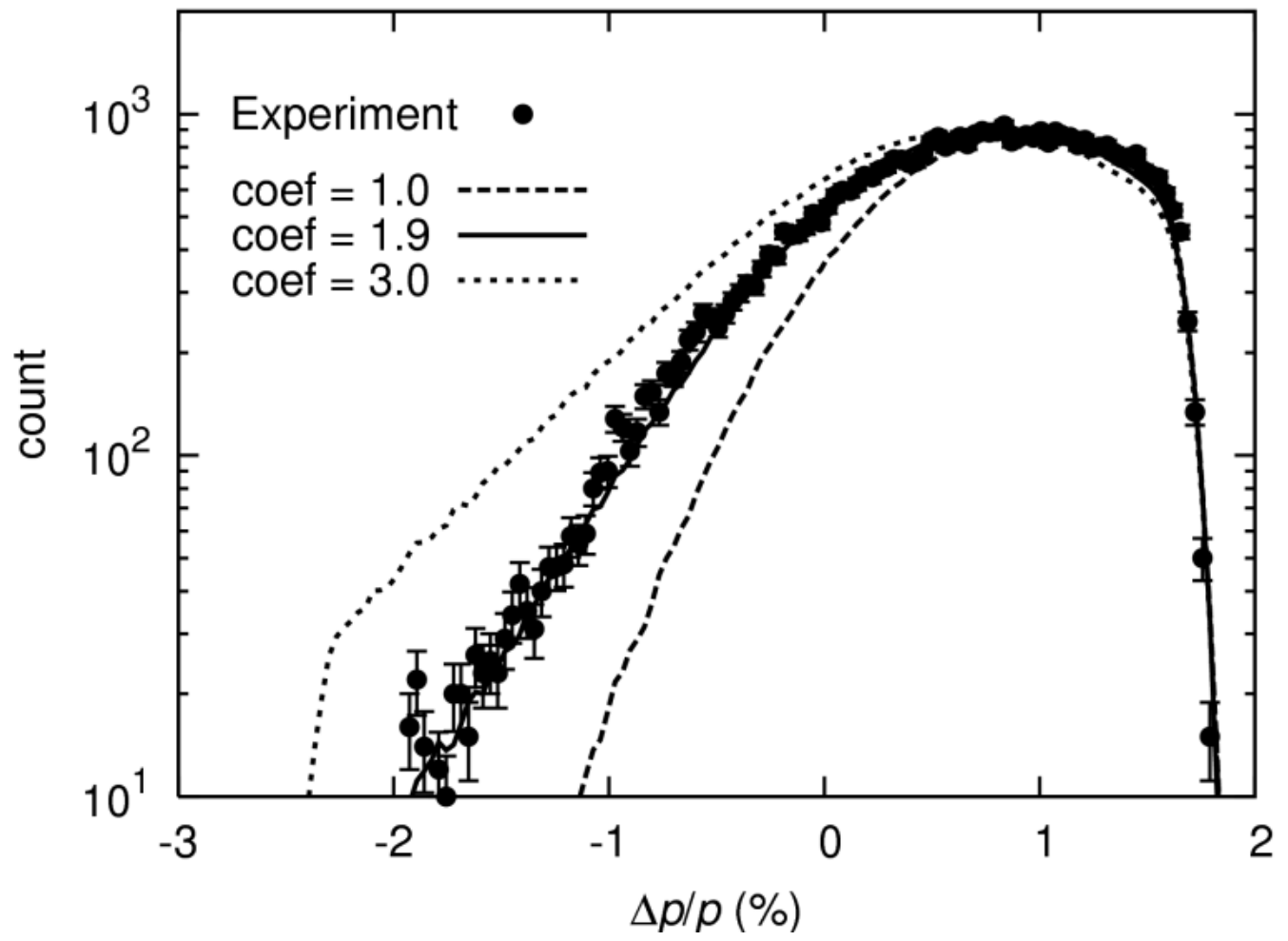

Fig. 3. Momentum distribution of $^{99}$Rh which are produced in the reaction $^{124}$Xe + Be (4.03 mm) at 345 MeV/nucleon. The experimental conditions are given as $^{97}$Pd +3% setting in Table 1. The dotted, solid, and dashed lines show the predictions from the LISE$^{++}$ simulations made in the Monte Carlo mode, in which the coef parameter is taken to be 3.0, 1.9, and 1.0, respectively. The simulation with coef = 1.9 gives the best agreement with the measurement. Note that the transmission of fragments on the high-momentum side is cut due to the F1 slit. See text.

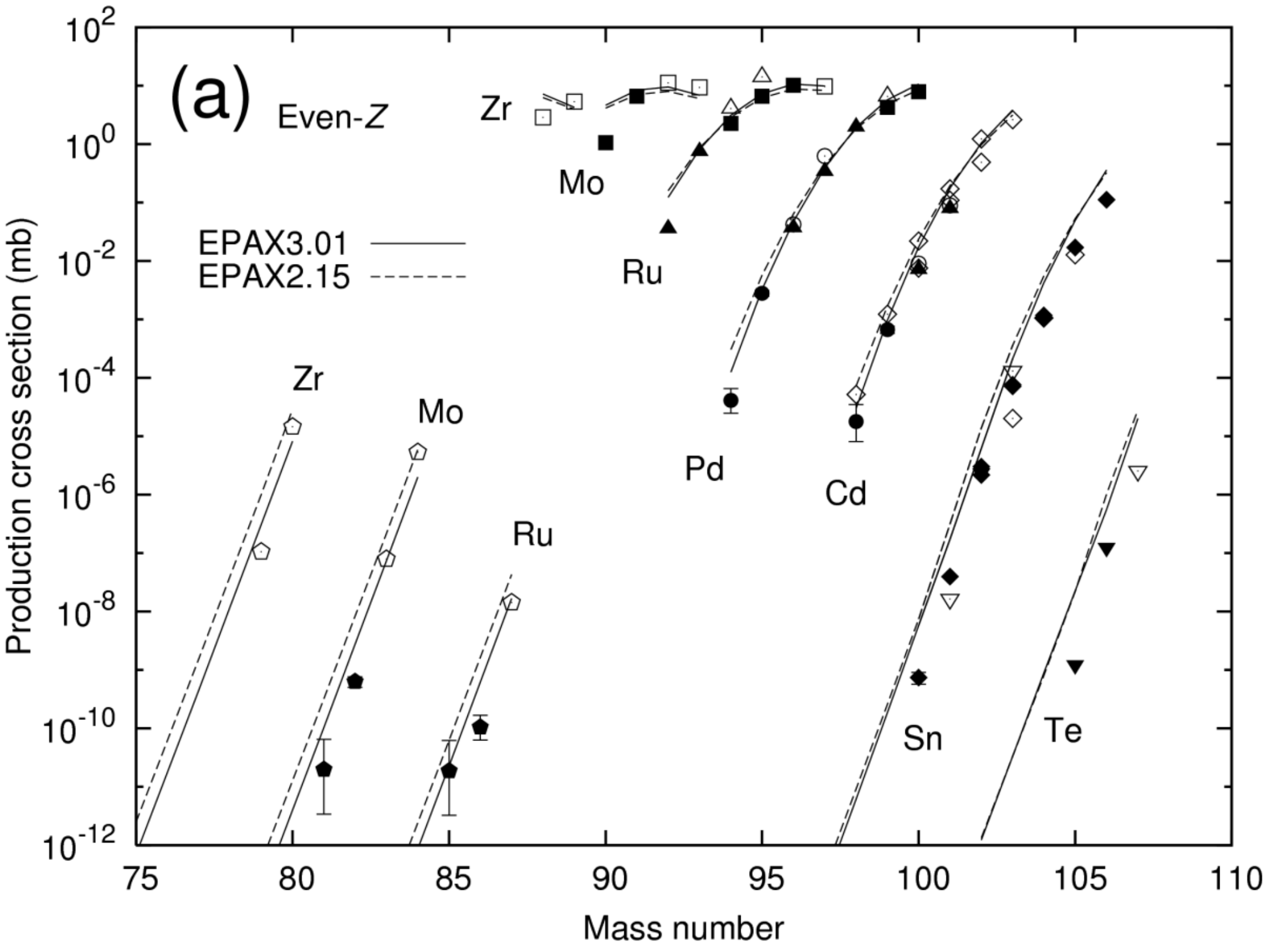

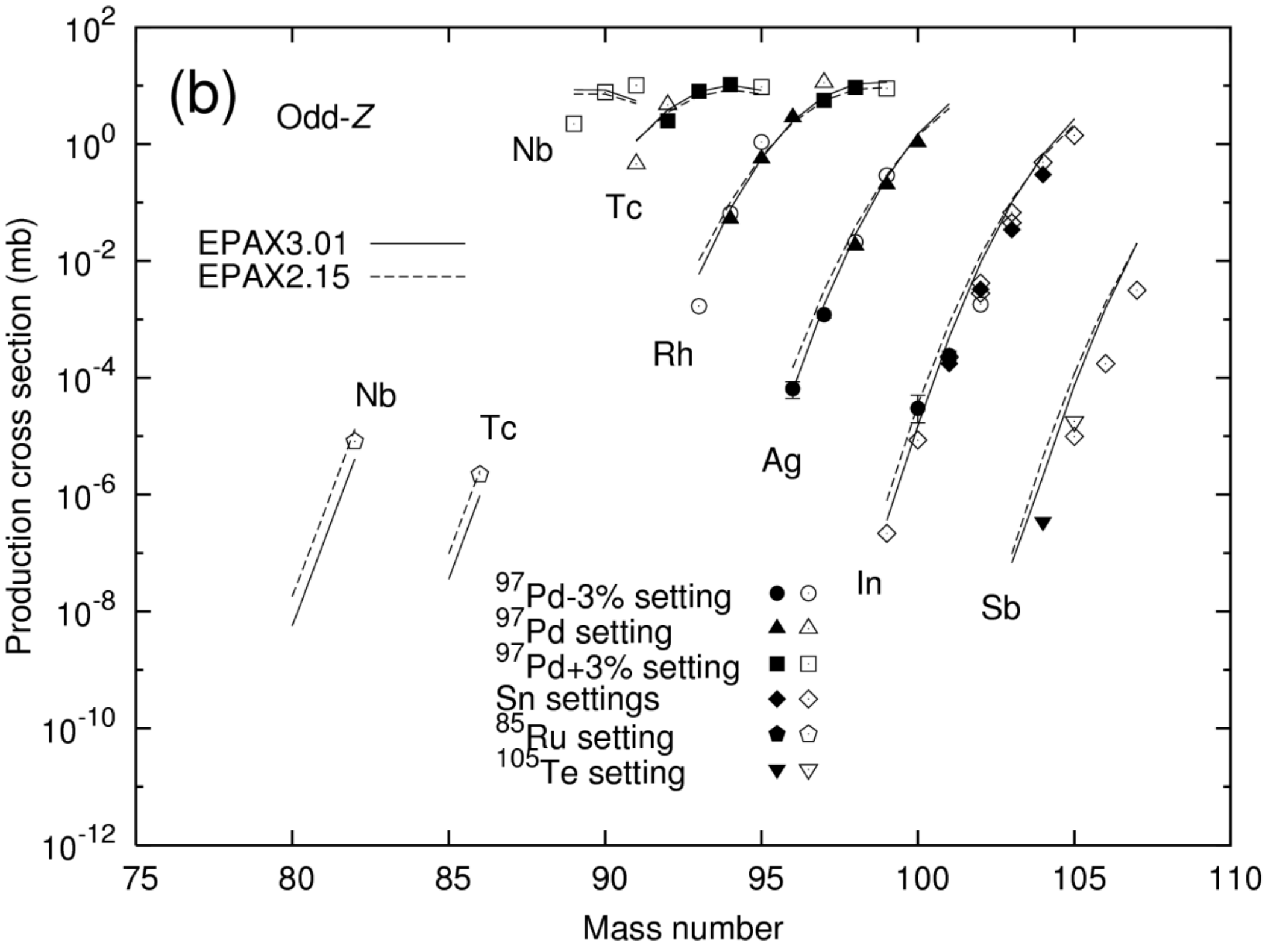

Fig. 4. Measured production cross sections of proton-rich isotopes produced in the reaction $^{124}Xe$ + Be (4.03 mm) at 345 MeV/nucleon shown along with predictions from the EPAX empirical cross-section formulae. (a) Results for even-$Z$ isotopes. (b) Results for odd-$Z$ isotopes. The circles, triangles, squares, diamonds, pentagons, and inverted triangles indicate the isotopes produced in the $^{97}$Pd –3%, $^{97}$Pd, $^{97}$Pd +3%, $^{100\text{-}105}$Sn, $^{85}$Ru, and $^{105}$Te settings, respectively (see Table 1). Filled symbols indicate that the distribution peak is located inside the slit opening at each focus and the open symbols indicate that it is located outside at some foci. Solid and dashed lines show the predictions from the EPAX3.01 and EPAX2.15 formulae, respectively. Errors shown are statistical only. See text.

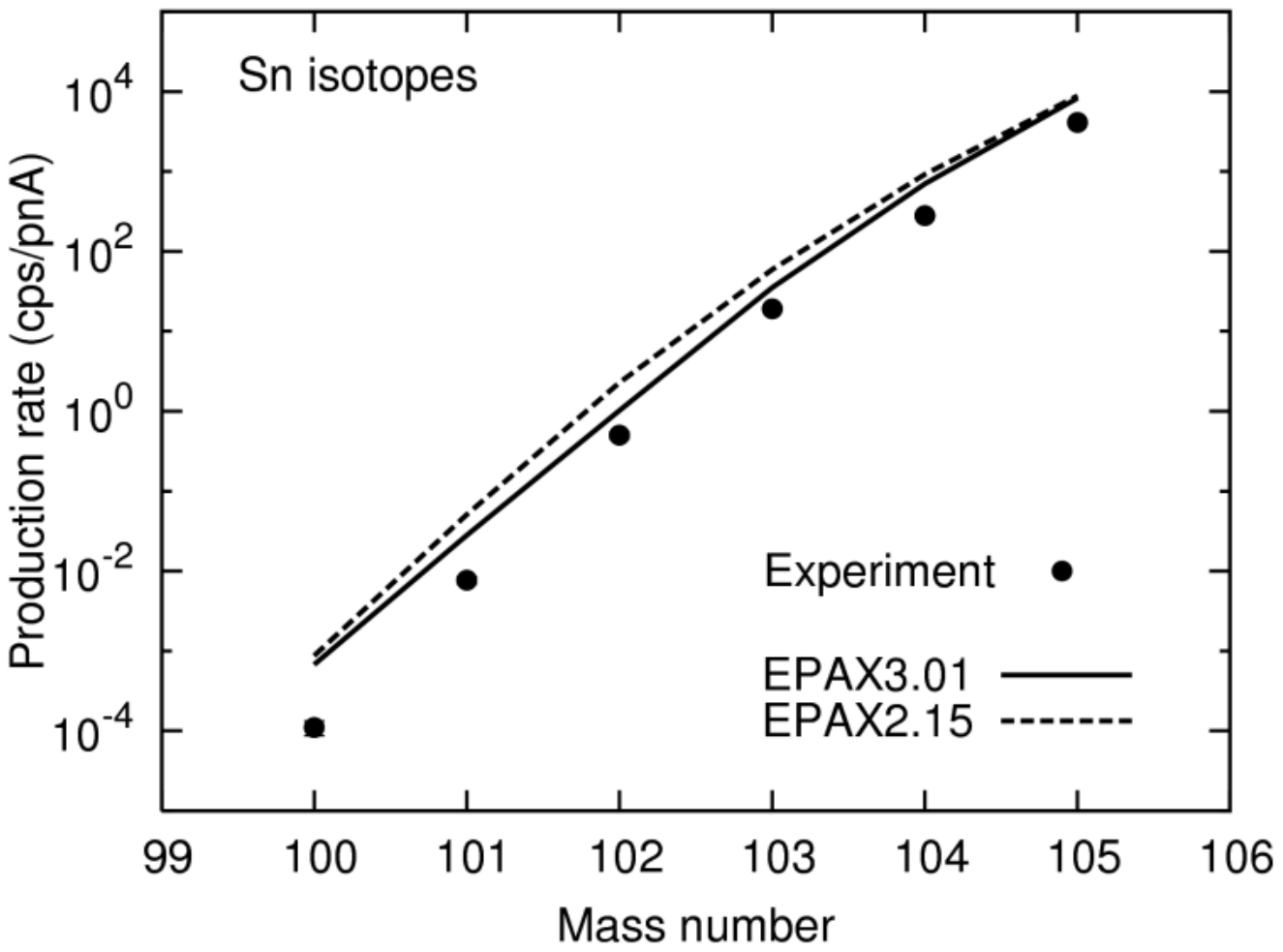

Fig. 5. Measured production rates of $^{100\text{-}105}$Sn isotopes produced in the reaction $^{124}$Xe + Be (4.03 mm) at 345 MeV/nucleon shown along with the LISE$^{++}$ simulations using the empirical cross-section formulae EPAX 3.01 (solid line) and EPAX2.15 (dashed line). The errors shown are statistical only. See text.

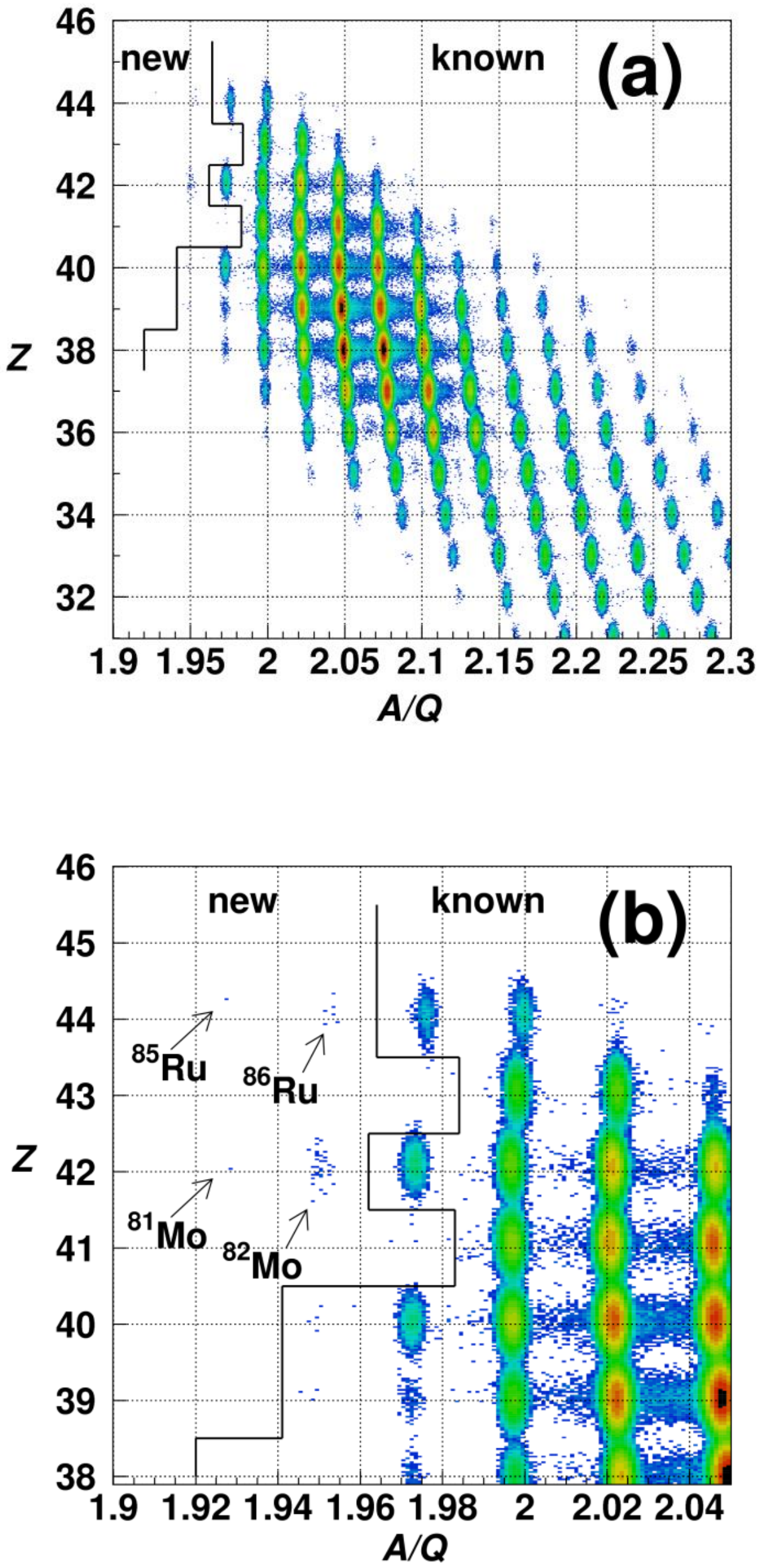

Fig. 6. (a) $Z$ versus $A/Q$ particle identification plot for projectile fragments produced in the reaction $^{124}Xe$ + Be (4.03 mm) at 345 MeV/nucleon. The experimental conditions are given as $^{85}Ru$ setting in Table 1. The limits of known isotopes are shown by solid lines. New isotopes are located on the left side of the solid lines. (b) Enlarged view for isotopes with $Z$ = 38 to 46. $^{85,86}Ru$ and $^{81,82}Mo$ are the new isotopes that we identified.

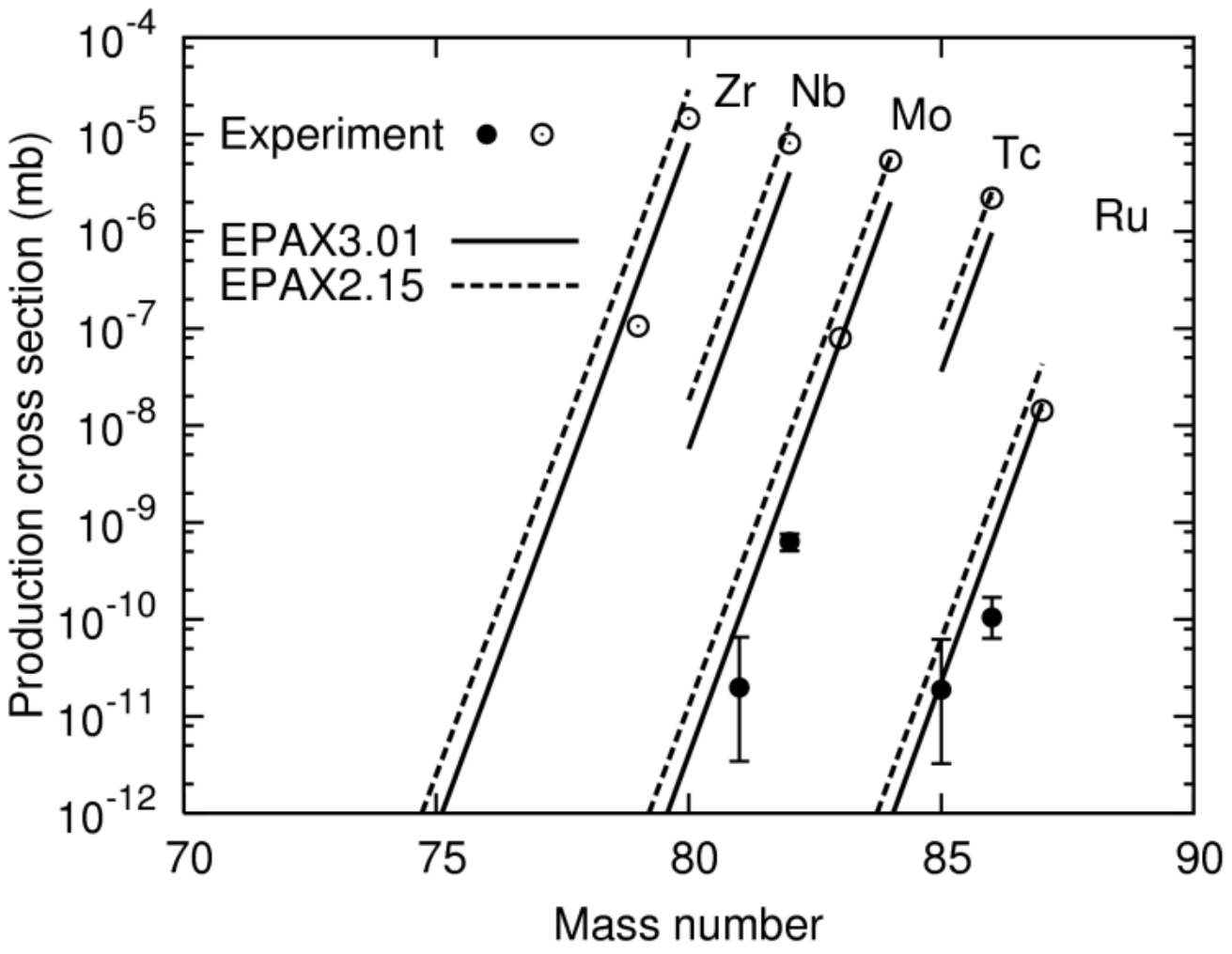

Fig. 7. Measured production cross sections of observed new isotopes and neighboring known isotopes shown along with the predictions from EPAX3.01 (solid lines) and EPAX2.15 formulae (dashed lines). The definition of the filled and open symbols is the same as given in in Fig. 4. Errors shown are statistical only.

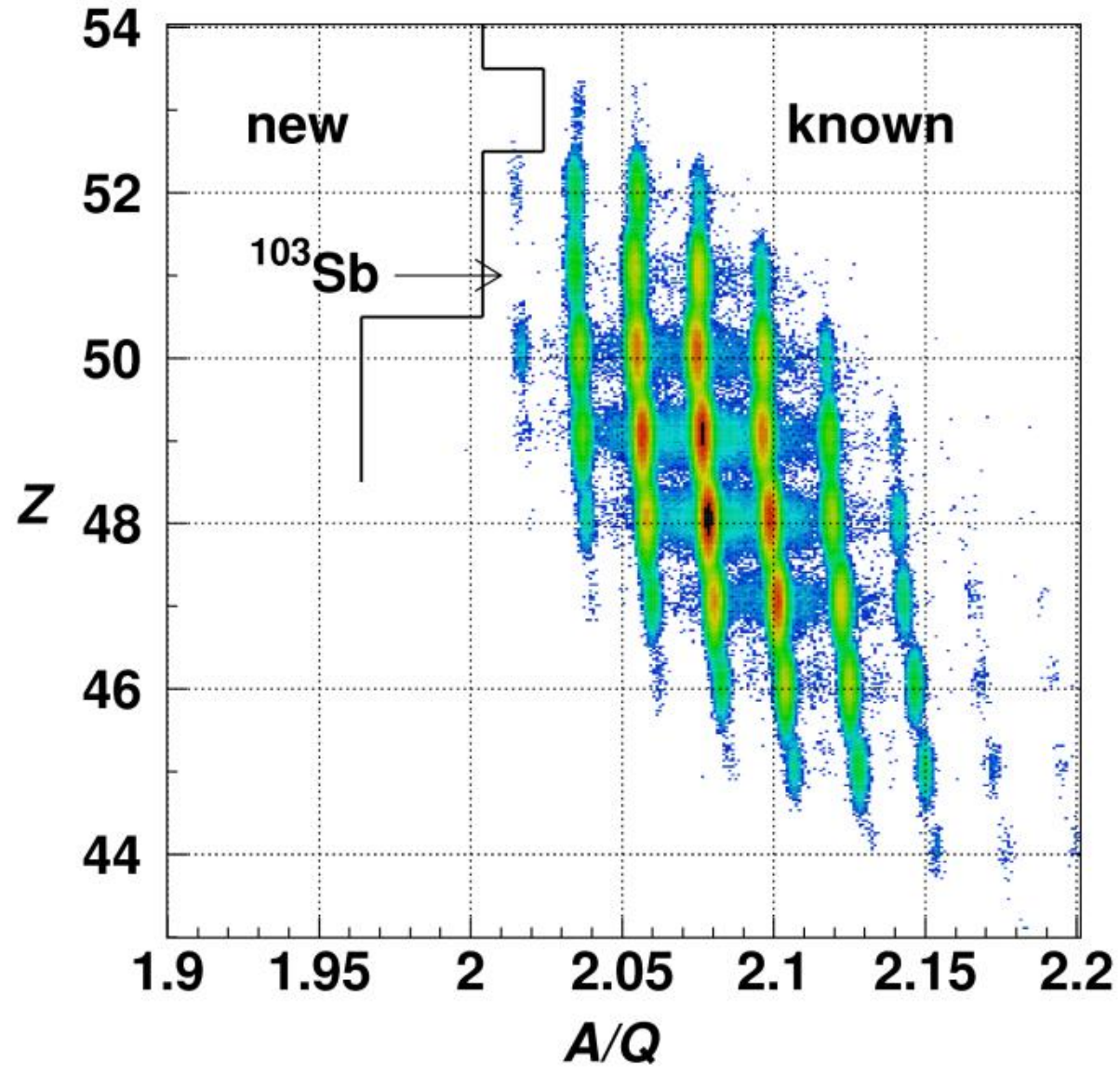

Fig. 8. $Z$ versus $A/Q$ particle identification plot for projectile fragments produced in the reaction $^{124}$Xe + Be (4.03 mm) at 345 MeV/nucleon. Experimental conditions are given as $^{105}$Te setting in Table 1. The known limits are indicated by solid lines.

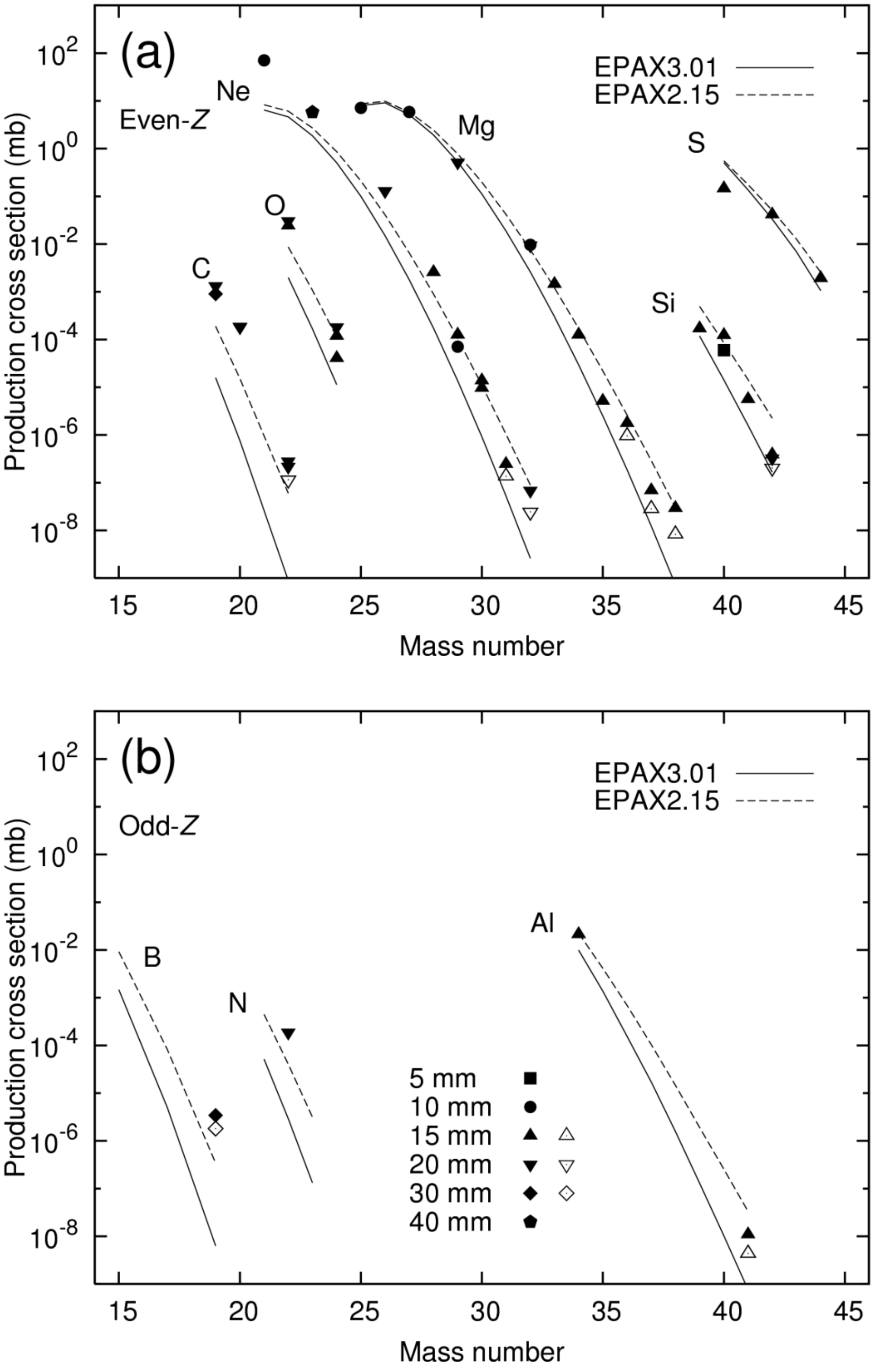

Fig. 9. Measured production cross sections of neutron-rich isotopes produced in the reaction $^{48}$Ca + Be at 345 MeV/nucleon shown along with the predictions from EPAX3.01 (solid lines) and EPAX2.15 formulae (dashed lines). (a) Results for even-$Z$ isotopes. (b) Results for odd-$Z$ isotopes. The squares, circles, triangles, inverted triangles, diamonds, and pentagons represent the experimental data obtained with the Be production targets with thicknesses of 5, 10, 15, 20, 30, and 40 mm, respectively. The errors shown are statistical only. We estimated that our method for determining the cross sections had a systematic error of ~50%. The open and filled symbols represent the cross sections with and without the correction for the effects of secondary reactions in the production targets, respectively. The corrected cross sections are shown only for the selected isotopes whose estimated augmentation factor is greater than ~2. See text.

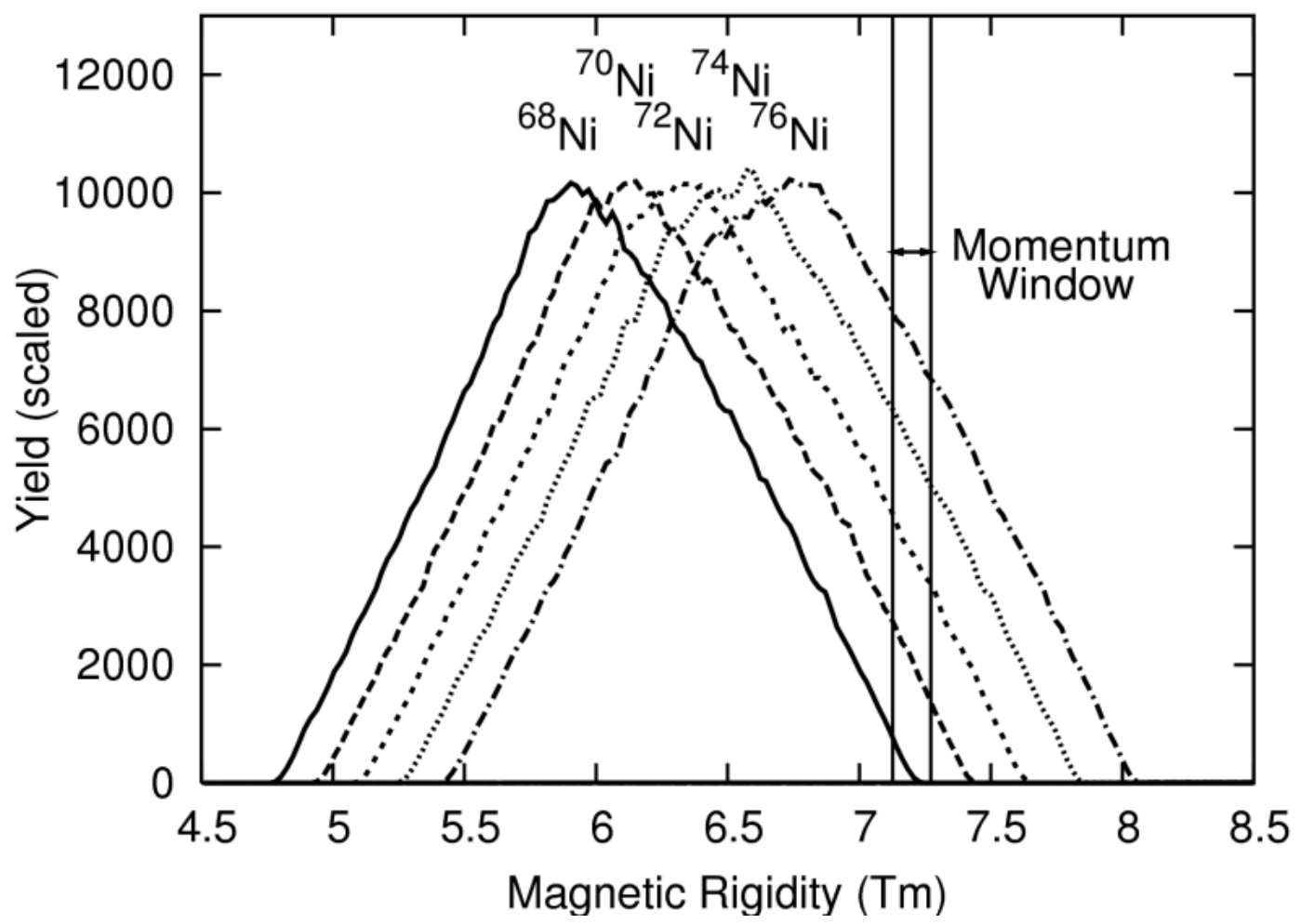

Fig. 10. Simulated $B\rho$ distribution of Ni isotopes produced in the reaction $^{238}U$ + Be (7 mm) at 345 MeV/nucleon. The $B\rho$ window of the BigRIPS separator, determined by the $B\rho$ setting and the slit opening at F1, are shown as momentum window (7.128 – 7.272 Tm) by the solid lines. The simulations were made using the LISE$^{++}$ code in Monte Carlo mode. The yields of the Ni isotopes are scaled.

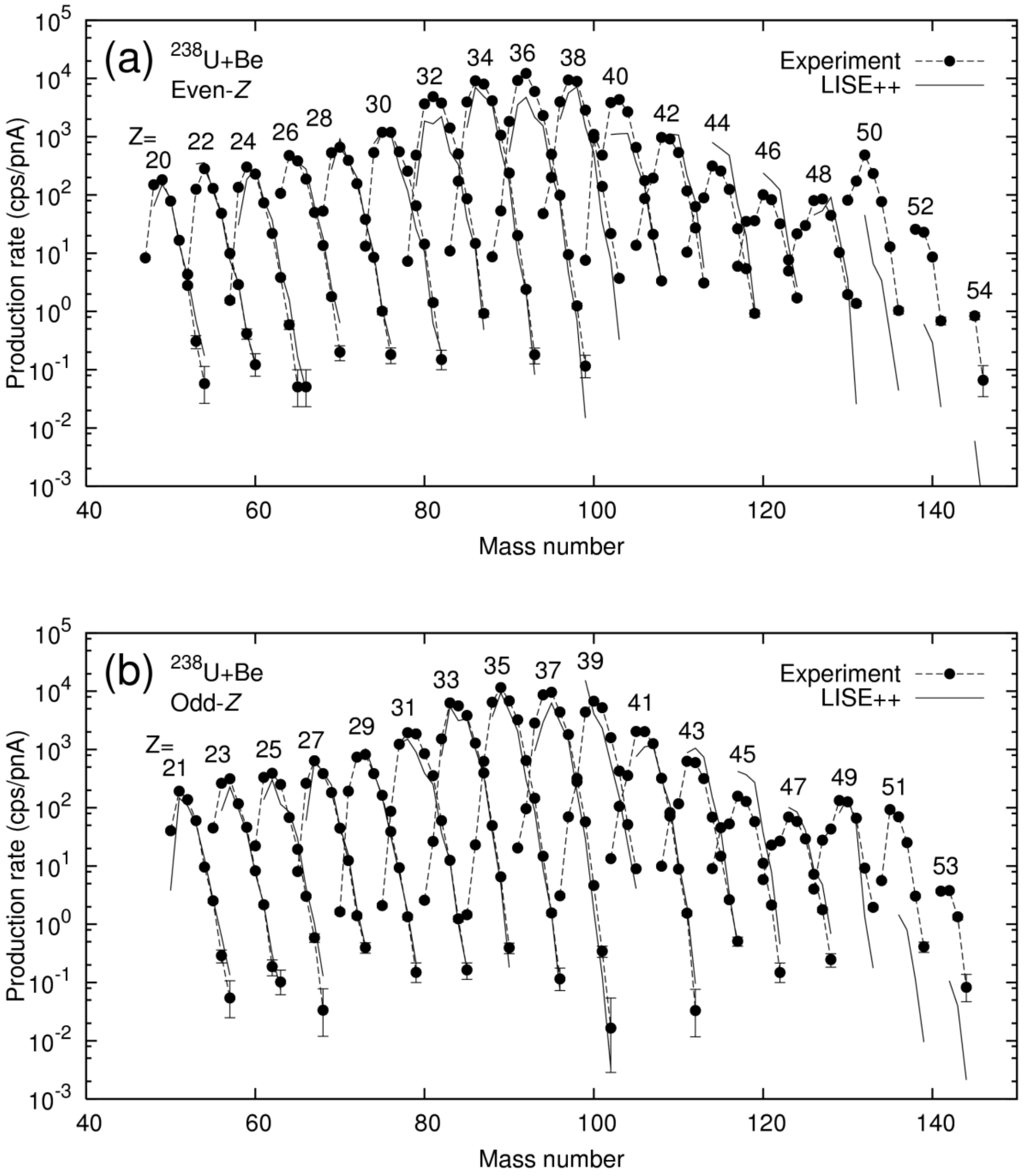

Fig. 11. Measured production rates for neutron-rich isotopes from the $^{238}U$ + Be reaction at 345 MeV/nucleon shown along with the predictions from the LISE$^{++}$ simulations. (a) Results for even-$Z$ isotopes. (b) Results for odd-$Z$ isotopes. The filled circles represent the experimental results, while the LISE$^{++}$ predictions are shown by the solid lines. The errors shown are statistical only. See text.

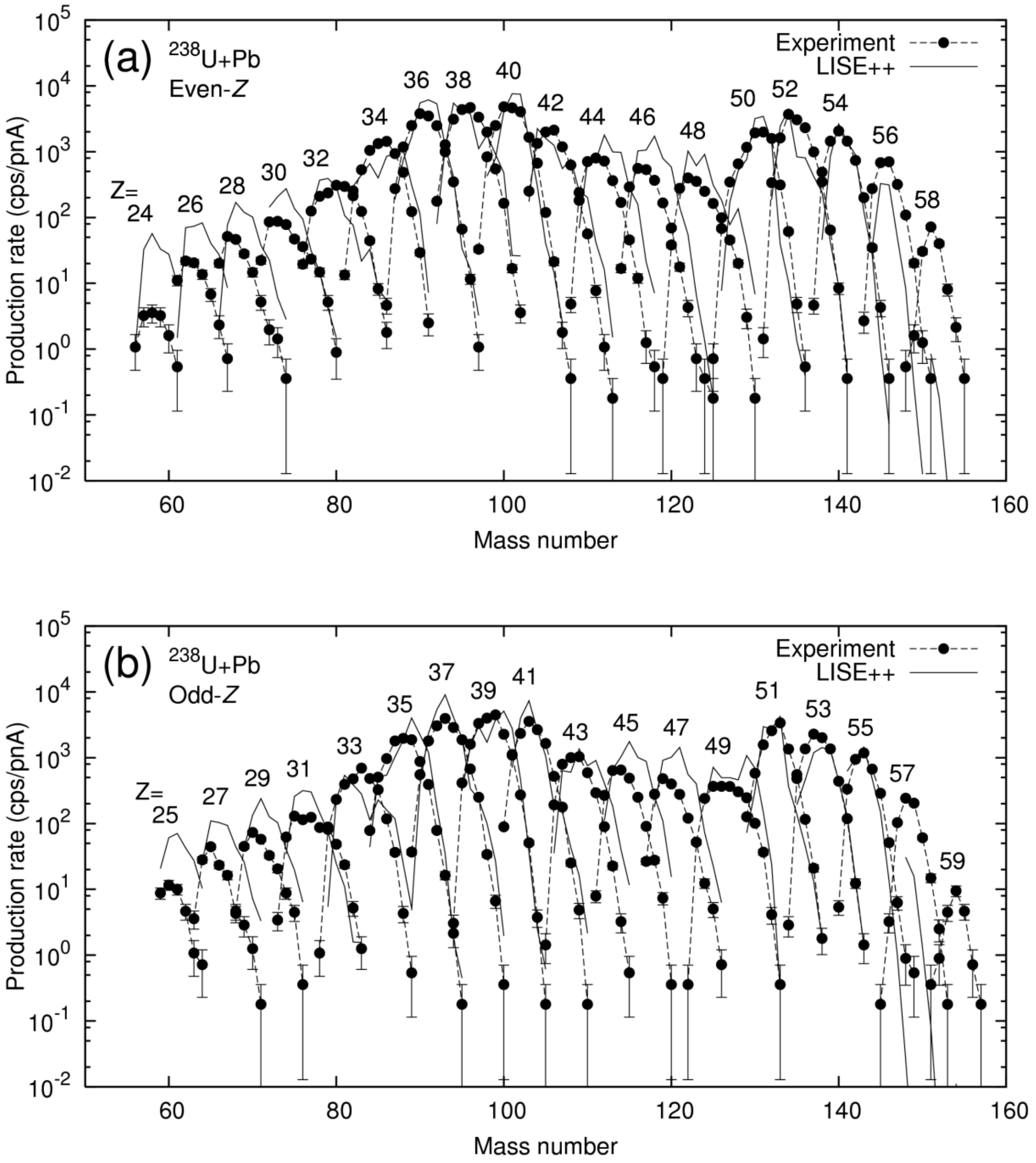

Fig. 12. Measured production rates for neutron-rich isotopes from the $^{238}$U + Pb reaction at 345 MeV/nucleon shown along with the predictions from the LISE$^{++}$ simulations. (a) Results for even-$Z$ isotopes. (b) Results for odd-$Z$ isotopes. The filled circles represent the experimental results, while the LISE$^{++}$ predictions are shown by the solid lines. The errors shown are statistical only. See text.